\documentclass{article}
\usepackage{arxiv}
\usepackage[utf8]{inputenc}
\usepackage[T1]{fontenc}
\usepackage{hyperref}
\usepackage{url}
\usepackage{booktabs}
\usepackage{amsfonts}
\usepackage{nicefrac}
\usepackage{microtype}
\usepackage{lipsum}
\usepackage{graphicx}
\usepackage{amsmath}
\usepackage{bm}
\usepackage{subcaption}
\usepackage{siunitx}
\usepackage{multirow}
\graphicspath{{./images/}}
\title{Global Mean-Amplitude Enhanced Spiking Neural Network Coherent Ising Machine}

\author{
Yan-Chen Jiang \\
State Key Laboratory of Information Photonics and Optical Communications\\
School of Physical Science and Technology\\
Beijing University of Posts and Telecommunications\\
Beijing 100876 China \\
  \texttt{yanchenjiang@bupt.edu.cn} \\
   \And
Lu Ma\\
MOT Key Laboratory of Transport Industry of Big Data Application Technologies for Comprehensive Transport\\
Beijing Jiaotong University\\
Beijing 100044 China\\
  \texttt{luma@bjtu.edu.cn} \\
  \And
Chuan Wang\\
The School of Artificial Intelligence\\
Beijing Normal University\\
Beijing 100875 China\\
  \texttt{chuanwang@bnu.edu.cn} \\
  \And
Tie-Jun Wang\\
State Key Laboratory of Information Photonics and Optical Communications\\
School of Physical Science and Technology\\
Beijing University of Posts and Telecommunications\\
Beijing 100876 China \\
  \texttt{wangtiejun@bupt.edu.cn} \\}

\begin{document}
\maketitle
\begin{abstract}
The coherent Ising machine (CIM) is a quantum-inspired computing platform that leverages optical parametric oscillation dynamics to solve combinatorial optimization problems by searching for the ground state of an Ising Hamiltonian. Conventional CIM implementations face challenges in handling non-uniform coupling strengths and maintaining amplitude stability during computation. In this paper, a new global mean-amplitude feedback-enhanced spiking neural network CIM (GFSNN-CIM) is introduced with a physics-driven amplitude stabilization mechanism to dynamically balance nonlinear gain saturation and coupling effects. This modification enhances synchronization in the optical pulse network, leading to more robust convergence under varying interaction strengths. Experimental validation on Max-Cut problems demonstrates that the GFSNN-CIM achieves up to a 27\% improvement in solution success rates compared to conventional spiking neural network CIM, with scalability improving as problem complexity increases. Further application to the traffic assignment problem (TAP) confirms the method's generality, the GFSNN-CIM achieves near-continuous accuracy (deviations $\textless$ 0.035\%) even at coarse discretization, while large-scale tests on Beijing’s road network (481 spins) validate its real-world applicability. These advances establish a physics-consistent optimization framework, where optical pulse dynamics directly encode combinatorial problems, paving the way for scalable, high-performance CIM implementations in complex optimization tasks.
\end{abstract}

\keywords{Quantum Squeezing\and Combinatorial optimization problems\and Coherent Ising machines\and Spiking neural networks\and Traffic assignment problem\and Route choices}

\section{Introduction}
The degenerate optical parametric oscillator (DOPO) based coherent Ising machine (CIM) \cite{wang2013coherent,inagaki2016coherent,pierangeli2019large,okawachi2020demonstration,honjo2021100,takesue2025finding} is a quantum-inspired computing scheme for solving combinatorial optimization problems (COPs), which is capable of room-temperature operation and scalability, offering advantages such as high speed and programmability. By utilizing quantum squeezing effects to manipulate coherent states in quantum optical systems, CIM establishes a dedicated computational framework that transforms COPs into energy minimization tasks solvable with photons. COPs suffer from exponential solution space expansion with growing variables, creating insurmountable computational burdens for traditional computing paradigms.  CIMs overcome this limitation through their unique physical implementation, achieving linear time-complexity scaling that offers transformative potential for large-scale optimization. A recent NTT study demonstrated that their CIM can handle problems with 100,000 spins, solving a 100,000-node Max-Cut problem \cite{honjo2021100} and a 40,000-node maximum independent set problem \cite{takesue2025finding}. However, CIMs face a fundamental challenge of amplitude heterogeneity in practical applications. This issue causes the system energy to freeze at excited-state levels of the Ising model, thereby preventing proper mapping to the ground state.  \cite{yamamoto2017coherent, leleu2019destabilization, kako2020coherent,inui2022control}. In existing research, the spiking neural network-based CIM (SNN-CIM) partially mitigates the heterogeneity of amplitude through dynamic dissipation parameter adjustment, yet retains inherent limitations in regulating strong coupling\cite{lu2022quantum,lu2023speed,lu2023combinatorial}. When real-world COPs are mapped to Ising models, prevalent coupling imbalance phenomena emerge, including orders-of-magnitude differences in coupling strength and localized strong coupling. These characteristics substantially degrade the solution quality of SNN-CIM. While in Ref.\cite{takesue2025finding}, the study also addresses the imbalance between external field and interaction terms, suggesting that non-uniform coupling relationships remain a challenge. 

Beyond solving mathematical problems like Max-Cut and Maximum Independent Set, the fundamental goal of CIM development is to address real-world combinatorial optimization challenges with practical significance. The transportation sector extensively involves a series of COPs, such as the Traveling Salesman Problem (TSP) \cite{lucas2014ising,warren2020solving},Vehicle Routing Problem (VRP) \cite{irie2019quantum,harikrishnakumar2020quantum,borowski2020new}, and traffic signal optimization \cite{inoue2021traffic,hussain2020optimal}. In transportation, quantum computing methods have been applied to diverse scenarios, including traffic assignment problem (TAP) \cite{neukart2017traffic,tambunan2023quantum,chitty2024applying}, signal optimization, traffic flow prediction \cite{thenmozhi2024quantum}, air traffic deconfliction \cite{stollenwerk2019quantum}. Among these, TAP has emerged as a challenges in quantum computing due to its complex network topology. In 2024, Chitty et al. proposed a quadratic unconstrained binary optimization (QUBO) formulation for TAP and validated its feasibility via quantum annealing, yet practical applications remain challenged by hardware immaturity and model oversimplifications \cite{chitty2024applying}. Notably, in 2017, Neukart's team at Volkswagen employed D-Wave's quantum annealing platform to explore traffic flow optimization using a hybrid quantum-classical approach \cite{neukart2017traffic}. This study neglected variations in factors such as segment length and road class. Although Tambunan (2022) introduced road segment weight parameters, the study was still confined to simplified networks and fixed-weight assumptions \cite{tambunan2023quantum}. These studies indicate that despite quantum computing's theoretical advantages, its practical deployment is constrained by hardware limitations (e.g., cryogenic requirements, limited qubit connectivity) and model simplifications.

In this paper, we proposes a global mean-amplitude feedback-enhanced SNN-CIM (GFSNN-CIM) that dynamically balances non-uniform coupling effects: suppressing amplitude divergence in strongly-coupled regions while enhancing cooperative search capability in weakly-coupled regions. Simulation experiments demonstrate that for Max-Cut problems with varying coupling strengths, GFSNN-CIM achieves higher optimal solution discovery probability than SNN-CIM, with this advantage becoming more pronounced as problem complexity increases. This study investigates the application of GFSNN-CIM to the TAP. The problem involves choosing vehicle routes from a set of origins to given destinations to alleviate traffic network congestion. When TAP incorporates route choice, the problem transforms into a COP. Compared with existing studies \cite{neukart2017traffic,tambunan2023quantum,chitty2024applying}, we formulate an Ising model for TAP incorporating flow discretization conditions, while comprehensively accounting for complex factors including multiple origin-destination (OD) pairs in real traffic scenarios. To validate the model effectiveness, we first conduct comparative verification against the classical Frank-Wolfe (FW) algorithm \cite{beckmann1956studies,frank1956algorithm} on a 5×5 grid network, followed by empirical performance evaluation of GFSNN-CIM on Beijing's real road network. Simulation results demonstrate that GFSNN-CIM yields superior computational outcomes compared to SNN-CIM for TAP.

The novel contributions of this paper are summarized as follows.
\begin{itemize}
    \item Proposal of a global mean-amplitude feedback-enhanced SNN-CIM that achieves up to 27\% higher success rate than SNN-CIM in solving Max-Cut problems.
    \item  Realization of Ising model mapping for the TAP under flow discretization conditions, validated on a 5×5 grid network (15 OD pairs, 75 vehicles): when using 1-vehicle discretization units (226 spins), GF-SNNCIM shows merely 0.0344\% deviation from the FW algorithm, with the discretization granularity refined to 0.1 vehicles (2,251 spins), this deviation reduces to 0.0099\%.
    \item  Experimental validation on Beijing's real road network (89 nodes, 408 links, 481 spins) confirms GFSNN-CIM's capability in handling complex conditions including multiple OD pairs, demonstrating the implementation potential of quantum-inspired algorithms in smart transportation systems.

\end{itemize}

\section{Methods}
\subsection{Global mean-amplitude feedback-enhanced coherent Ising machine based on spiking
neural networks}
The Ising model \cite{ising1924beitrag} characterizes binary competition phenomena in many-body systems, such as particle spins. Spin is an intrinsic quantized property of particles, which can only occupy two discrete states (e.g.,  1 or -1; \textuparrow or \textdownarrow). The global system behavior is governed by competitive interactions between spin states. For an Ising system comprising N spins, its Hamiltonian is expressed as \(H=-{\sum_{ij}J_{ij}}\sigma_{i}\sigma_{j}\), where $\sigma_{i(j)}\in \{ 1,-1 \}$, and \(J_{ij}\) represents the coupling strength between spins \(\sigma_i\) and spin \(\sigma_{j}\). The CIM, leveraging quantum optical squeezing effects, encodes spin systems into nonclassical optical fields to achieve ground-state solutions \cite{wang2013coherent,wang2015coherent}. Its core component is the DOPO, where each DOPO unit corresponds to an Ising spin. Within the DOPO, a nonlinear crystal converts input pump photons into signal and idler photons with identical frequencies. This parametric down-conversion process generates nonclassically correlated states. In the optical resonator, signal light amplification is phase-sensitive relative to the pump field. Maximum gain occurs at relative phases of 0 or \(\pi\). The phase bistability of DOPO outputs directly maps onto spin 1 or -1 states, establishing a photonic-qubit representation. Compared to classical algorithms, the CIM harnesses quantum coherence of optical fields for parallel state space exploration. Its nonlinear gain dynamics effectively circumvents local optima, enabling rapid convergence to the Hamiltonian ground state.

An CIM system with N DOPOs exhibits a preference for stabilizing at the position of minimal total photon decay rate, which is proportional to the Hamiltonian of the Ising problem \cite{wang2013coherent,wang2015coherent}. The total photon decay rate \(\Gamma\) is
\begin{eqnarray}
\Gamma=N-{\sum _{1 \leq i<j \leq N} J _{ij}}\sigma_{i}\sigma_{j}+O(\frac{\epsilon^3N^4}{(P-1)^3}),
\end{eqnarray}
where \(\epsilon\) is \(max|J_{ij}|\), and \(P\) is the intensity of the pump light. Higher-order correction terms can be evaluated when couplings of the same strength exist between any two of the DOPOs\cite{wang2013coherent,wang2015coherent} . In a specific phase configuration of the DOPOs, the distinction in the total photon decay rates between scenarios involving mutual coupling and those lacking it precisely corresponds to the energy of the respective spin configuration in an Ising problem. A global mode that achieves the minimum \(\Gamma\) serves as the minimum Hamiltonian solution. This working principle of CIM involves progressively increasing the pump intensity to induce competition among global modes, until the DOPOs oscillate in the state with the minimal total photon decay rate, thereby identifying the global mode corresponding to the solution to the Ising problem. This principle is known as the minimum gain principle.

In large-scale CIMs, amplitude heterogeneity causes the system energy to freeze at excited states of the Ising model, preventing proper mapping to the ground state. This challenge can be addressed through the integration of nonlinear optical spiking neural networks\cite{lu2023combinatorial,lu2023speed}. By tuning the dissipative parameters of spiking neurons, constructed using antisymmetrically coupled DOPO and dissipative pulses, one can control the rate and direction of contraction in phase-space volumes. This mechanism effectively destabilizes metastable states, enabling the SNN-CIM to overcome local energy minima, as validated on Max-Cut problems with up to 2,000 nodes. \cite{lu2023combinatorial}. 

However, Ising problems converted from real-world COPs typically exhibit significant non-uniform coupling characteristics (including orders-of-magnitude differences in coupling strength and localized strong coupling), exposing limitations in the local parameter adjustment mechanism of conventional SNN-CIM. So, we proposed GFSNN-CIM introduces global mean-amplitude feedback, which generates a normalized feedback factor $\zeta\frac{\sum_j{x_j}}{\sqrt{N} }$ by monitoring of the collective oscillatory states of spiking neurons. This normalized factor modulates the evolution of in-phase amplitudes through two distinct mechanisms. In strongly coupled regions, it suppresses amplitude overshooting via negative feedback to prevent energy landscape distortion. In weakly coupled regions, it enhances signal synchronization through positive feedback to increase the probability of escaping local minima.

Based on this principle, the dynamics of the DOPO network with $N$ spiking neurons can be described by 
\begin{equation}
\begin{aligned}
\frac{dx_i}{dt} &= a x_i - x_i^3 + J_{xk} b k_i + I + \zeta\frac{\sum_j{x_j}}{\sqrt{N} }\\
\frac{dk_i}{dt} &= -b k_i + J_{kx} x_i, \\
I &= \tanh\left(c\sum_j J_{ij} x_j\right),
\end{aligned}
\end{equation}
where \(x_i\) and \(k_i\) are the in-phase and dissipative pulse amplitudes. $a$ and $b$ denote the nonlinear gain generated and inherent dissipation of the system. $I$ denotes the external bias current stimulus, $c$ characterizes the interaction strength. $\zeta$ is the global mean-amplitude feedback coefficient. This feedback regulation maintains the dissipative advantages of SNN-CIM while enhancing the system's ground-state convergence probability. To systematically evaluate GFSNN-CIM's effectiveness, this paper employed numerical modeling to compare its ground-state convergence probability with SNN-CIM on Max-Cut problems of identical scale but with distinct edge-weight distributions.

\subsection{Ising model transformation for traffic assignment problem}
As a core task in intelligent transportation systems, the TAP commonly uses user equilibrium (UE) assumption to describe the behavior of travelers \cite{wardrop1952road,ortuzar2002modelling,perederieieva2015framework,chakroborty2017principles}.  UE mirrors the natural tendency of travelers to select path with the shortest time to reach their desired destination. TAP's mathematical model was established by Beckmann in 1956 \cite{beckmann1956studies,daskin1985urban}. A transportation system can be represented as a directed graph $G(A,B)$, where $A$ represents the set of directed links and $B$ denotes the set of nodes. Users travel from their origins to destinations, and the demand for each OD pairs $p\in Z$ is given by $D_p$, where $Z$ is the set of OD pairs. To account for traffic congestion, link cost functions are introduced into the
model. Let $t_a(f_a)$ denote a link
cost function of the link $a$ that depends on the link flows $f_a$, which is generally the BPR function  (\(t_a(f_a)=t_{a,0}[1+\alpha(\frac{f_a}{f_{a,{\text{capacity}}}})^\beta]\)) \cite{maerivoet2005transportation}. Here, \(t_{a,0}\) and $f_{a,{\text{capacity}}}$ represent the freeFlow time and the capacity of the link $a$. The function parameters \(\alpha\) and \(\beta\) are generally taken as  0.15 and 4. If the flow of the path is $F_k$, so $f_a = \sum_{p\in{Z}}\sum_{k\in{K_p}}\delta_a^kF_k$. Here, \(\delta_a^k\) is the indicator function of the link $a$ and the route $k$, $k\in K_p$ is the path for the OD pair p. The mathematical model for UE condition by minimizing the cost function\cite{daskin1985urban,patriksson2015traffic,bertsekas1998network}
\begin{equation}
\begin{aligned}
\min&{\sum _{a\in{A}}\int_{0}^{f_a}t_a(x)dx}, ~~~~~~~s.t.&
\left\{
\begin{aligned}
    &\sum _{k\in{K_p}}F_k = D_p,    \forall p \in Z,\\&F_k \geq 0,    \forall k \in K_p,    \forall p \in Z,\\&f_a = \sum_{p \in Z} \sum_{k \in K_p} \delta_a^k F_k,    \forall a \in A.
\end{aligned}
\right.
\end{aligned}
\end{equation}
The most famous algorithm for solving Eq. (3) was FW algorithm, a link-based algorithm for convex
optimization problems.  Although FW cannot guide real-world travelers in designing travel routes, based on its wide representativeness, this article uses it to explore this problem. Furthermore, the Dijkstra-based incremental assignment (DIA) algorithm, another classical algorithm for TAP, can result in a solution that closely matches the UE condition and also is a algorithm we applied.

The following text introduces the mapping of the TAP to the Ising model. The travelers across all OD pairs are discretized into $N$ vehicle groups, with each group containing $g$ vehicles. Each vehicle group $i$ ($i\in[1,N]$) is assigned $M$ alternative routes. Considering computational feasibility and industry practices in commercial navigation systems (which typically offer 3 route options), we set $M=3$ in this study. A binary decision variable $q_{ij}\in\{0,1\}$ is defined to indicate route selection, where $q_{ij}=1$ denotes that vehicle group $i$ chooses alternative route $j$, and $q_{ij}=0$ otherwise. The binary variables $q_{ij}$ establish a mathematical representation of link flows through their mapping relationship with route choice decisions. Considering the initial link flow $f^0_a$, the total flow on link $a$ is \(f_a=g\cdot \sum_{ij}q_{ij}\delta_{ij,a}+f^0_a \). The cost function can be expressed as
\begin{equation}
\begin{aligned}
&\min{\sum _{a\in A}\int_{0}^{f_a}t_a(x)dx}
\\&=\min{\sum _{a\in A}\int_{0}^{f_a}t_{a,0}[1+\alpha(\frac{x}{f_{a,{\text{capacity}}}})^\beta]dx}
\\&=\min{\sum _{a\in A}t_{a,0}[f_a+\frac{\alpha}{(\beta+1)({f_{a,{\text{capacity}}}})^\beta}f_a^{\beta+1}]}
\\&=\min{\sum _{a\in A}t_{a,0}[(g\cdot \sum_{ij}q_{ij}\delta_{ij,a}+f^0_a)+\frac{3}{100({f_{a,{\text{capacity}}}})^4}(g\cdot \sum_{ij}q_{ij}\delta_{ij,a}+f^0_a)^{5}]}.
\end{aligned}
\end{equation}

When using CIM to solve, each vehicle group is equipped with $M$ DOPO pulses to represent the selection of $M$ alternative routes. The measured value $Q_{ij}$ of phase 0 (\(\pi\)) of each DOPO pulse is +1 (-1). \(Q
_{ij}=2*q_{ij}-1\) (\(q_{ij}=\frac{Q_{ij}+1}{2}\) ) indicates whether the corresponding route is selected (\(Q_{ij}=1\)) or not selected (\(Q_{ij}=-1\)). Thus, the TAP can transform the cost function into the Ising model through \(f_a=g\cdot \sum_{ij}(\frac{Q_{ij}+1}{2})\delta_{ij,a}+f^0_a \). Additionally, each group can only choose one route from the alternative routes, leading to the constraint condition
\begin{equation}
\min~\lambda\sum_i^N(\sum_j^Mq_{ij}-1)^2,
\end{equation}
where \(\lambda\) represents the constraint coefficient. Therefore, the objective function of our scheme is $Obj(Q_{ij})$, 
\begin{equation}
\begin{aligned}
min&~Obj(Q_{ij})\\=&\min~\big\{{\sum _{a\in A}t_{a,0}\{[g\cdot \sum_{ij}q_{ij}\delta_{ij,a}+f^0_a]+\frac{3}{100({f_{a,{\text{capacity}}}})^4}[g\cdot \sum_{ij}q_{ij}\delta_{ij,a}+f^0_a]^{5}}\}\\&+\lambda\sum_i^N(\sum_j^Mq_{ij}-1)^2 \big\}
\\=&\min~\big\{{\sum _{a\in A}t_{a,0}\{[g\cdot \sum_{ij}(\frac{Q_{ij}+1}{2})\delta_{ij,a}+f^0_a]+\frac{3}{100({f_{a,{\text{capacity}}}})^4}[g\cdot \sum_{ij}(\frac{Q_{ij}+1}{2})\delta_{ij,a}+f^0_a]^{5}}\}
\\&+\lambda\sum_i^N[\sum_j^M(\frac{Q_{ij}+1}{2})-1]^2\big\}.
\end{aligned}
\end{equation}
When the constraint condition is satisfied, the value of Eq. (5) is 0. The solution of Eq. (6) is equivalent to Eq. (4)
\begin{equation}
\begin{aligned}
\min~Obj(Q_{ij})=\min~{\sum _{a\in A}\int_{0}^{f_a}t_a(x)dx}.
\end{aligned}
\end{equation}

Since the CIM requires the objective function to be quadratic, $Obj(Q_{ij})$ may contain up to quintic terms upon expansion. These higher-order terms must be reduced to quadratic form. We adopt a quadratic function approximation method to transform the objective function into a quadratic form, ensuring compatibility with the CIM framework (For example, $x +\frac{3}{100(25)^4}x^5$ can be effectively approximated by the quadratic polynomial $0.001899x^2 + 0.969102x + 0.128399$ within the interval $[8,10]$.) The quadratic approximation of the objective function is
\begin{equation}
\begin{aligned}
\min&~Obj(Q_{ij})\\\approx &\min~\big\{\sum_{a\in A} t_{a,0}\{\gamma_{a,1}[g\cdot \sum_{ij}(\frac{Q_{ij}+1}{2})\delta_{ij,a}+f^0_a]^2+\gamma_{a,2}[g\cdot \sum_{ij}(\frac{Q_{ij}+1}{2})\delta_{ij,a}+f^0_a]+\gamma_{a,3}\}\\&+\lambda\sum_i^N(\sum_j^M\frac{Q_{ij}+1}{2}-1)^2\big\}\\=&\min~(Q^TXQ+Q^TY+C)\\=&\min~Q^{\prime~T}JQ^{\prime}.
\end{aligned}
\end{equation}
Here, \(\gamma_{a,1}\), \(\gamma_{a,2}\), and \(\gamma_{a,3}\) are the parameters of the quadratic approximation function for link $a$, and are related to the flow variation range of the link. By expanding the quadratic approximation function, it can be reformulated as the polynomial expression of \(Q_{ij}\), which includes the quadratic term \(Q^TXQ\), linear term \(Q^TY\), and the constant term \(C\). \textbf{$Q=[Q_{11}, Q_{12}, ..., Q_{NM}]$} is the vector of $Q_{ij}$. $X$ and $Y$ are matrices and vectors that store correlation coefficients. Then the quadratic term is retained and the linear term is transformed into a new quadratic term by adding an auxiliary pulse \(Q_{aux}\), and the constant term will be discarded. The measurement value of the auxiliary pulse is restricted to 1. In the end, the quadratic approximation function became \(Q^{\prime~T}JQ^{\prime}\), which is equivalent to the form of Ising Hamiltonian. \textbf{$Q^{\prime}=[Q_{11}, Q_{12}, ..., Q_{NM},Q_{aux}]$}, and \(J\) is a known coefficient matrix. Thus, the CIM can be employed to solve TAP incorporates route choices, with the required number of spins being \(N*M+1\).

 In the small scale TAP, the parameters of quadratic function on different links are the same. But in large-scale problems, the parameters on different links are different, which means that multiple quadratic functions are used. When using multiple quadratic functions, such as in the TAP task based on Beijing's urban road network, the solving process is divided into two steps: (1) An identical approximation quadratic function is used for all links,  and by using the CIM provides predicted flow estimates of all links. (2) Based on the predicted flows, the more accurate approximate quadratic function parameters can be determined for each link, leading to better solution results by CIM to solve the task again. This two-step approach balances computational efficiency and accuracy.

\section{Results}
In this section, we present the results of Max-Cut problem and TAP tasks (the 5\(\times\)5 grid network and Beijing urban road network) solved using the GFSNN-CIM. The device processor utilized is an AMD Ryzen 9 7945HX with Radeon graphics (2.50 GHz), and the graphics card is NVIDIA GeForce RTX 4060 laptop GPU (8 GB).

\subsection{The results of Max-Cut problem}

\begin{figure}[htbp]
    \centering
    \begin{subfigure}[t]{0.32\textwidth}
        \centering
        \includegraphics[width=\linewidth]{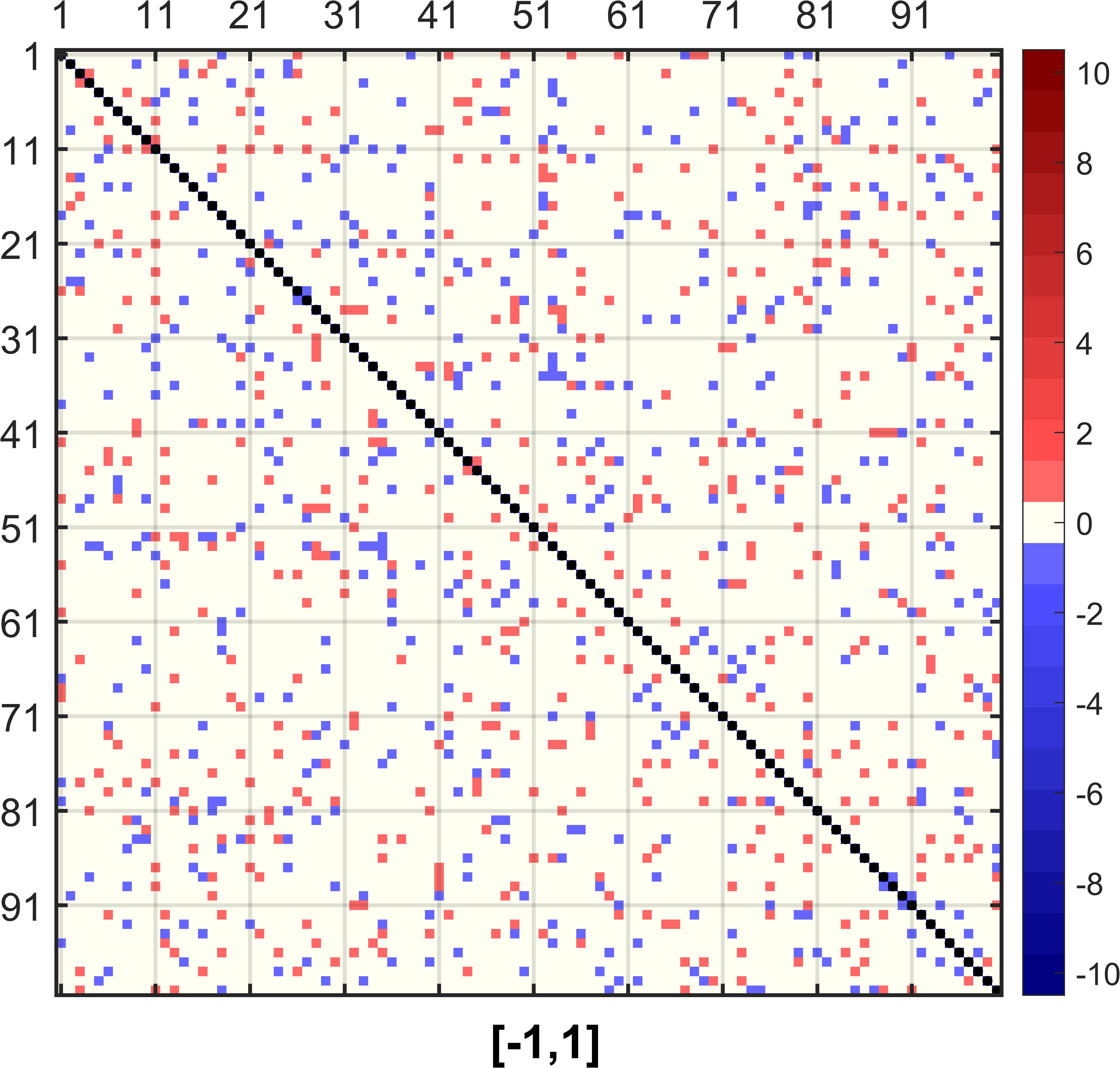}
        \caption{pm1s\_100.5}
    \end{subfigure}
    \hfill
    \begin{subfigure}[t]{0.32\textwidth}
        \centering
        \includegraphics[width=\linewidth]{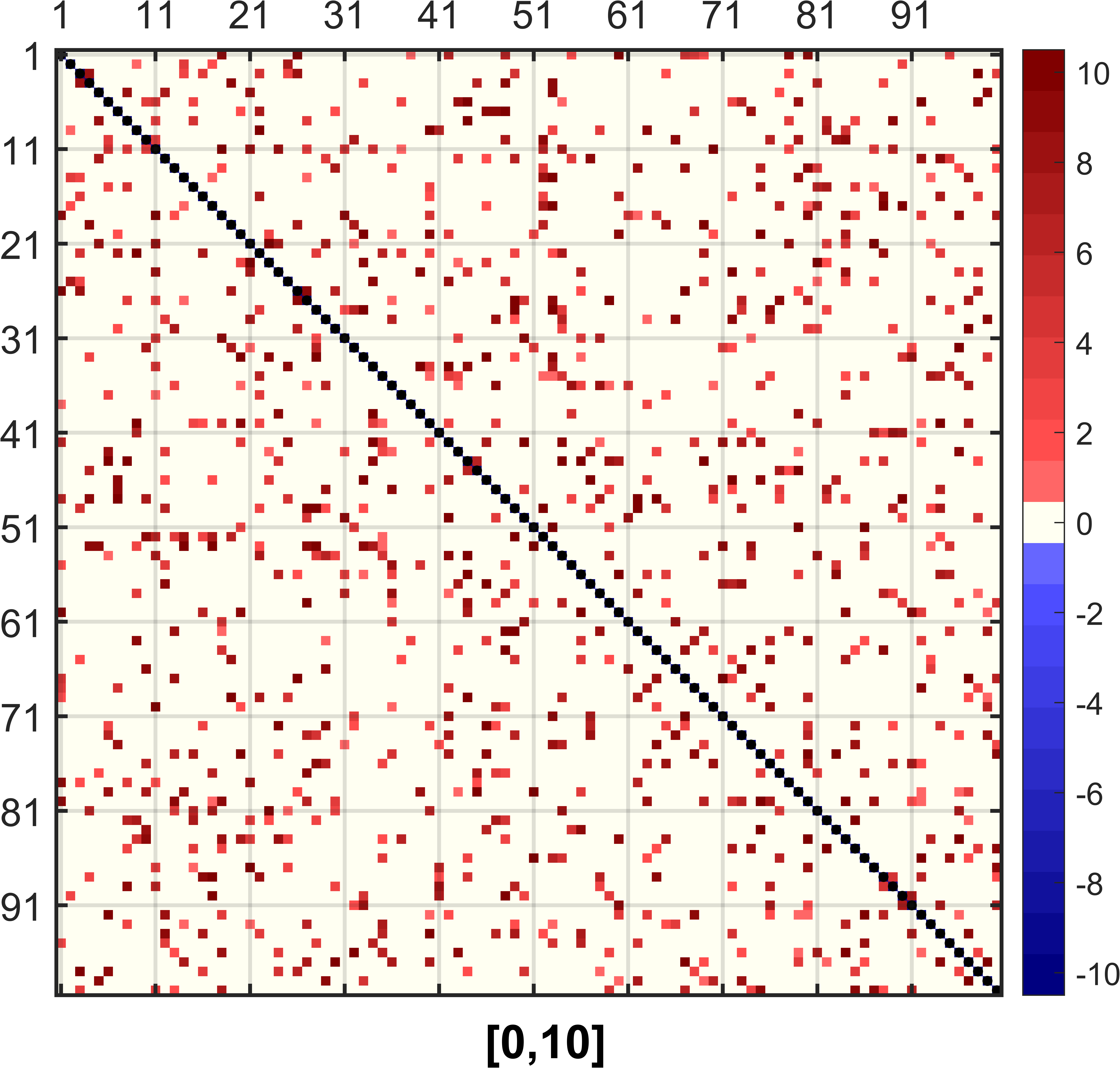}
        \caption{pw01\_100.5}
    \end{subfigure}
    \hfill
    \begin{subfigure}[t]{0.32\textwidth}
        \centering
        \includegraphics[width=\linewidth]{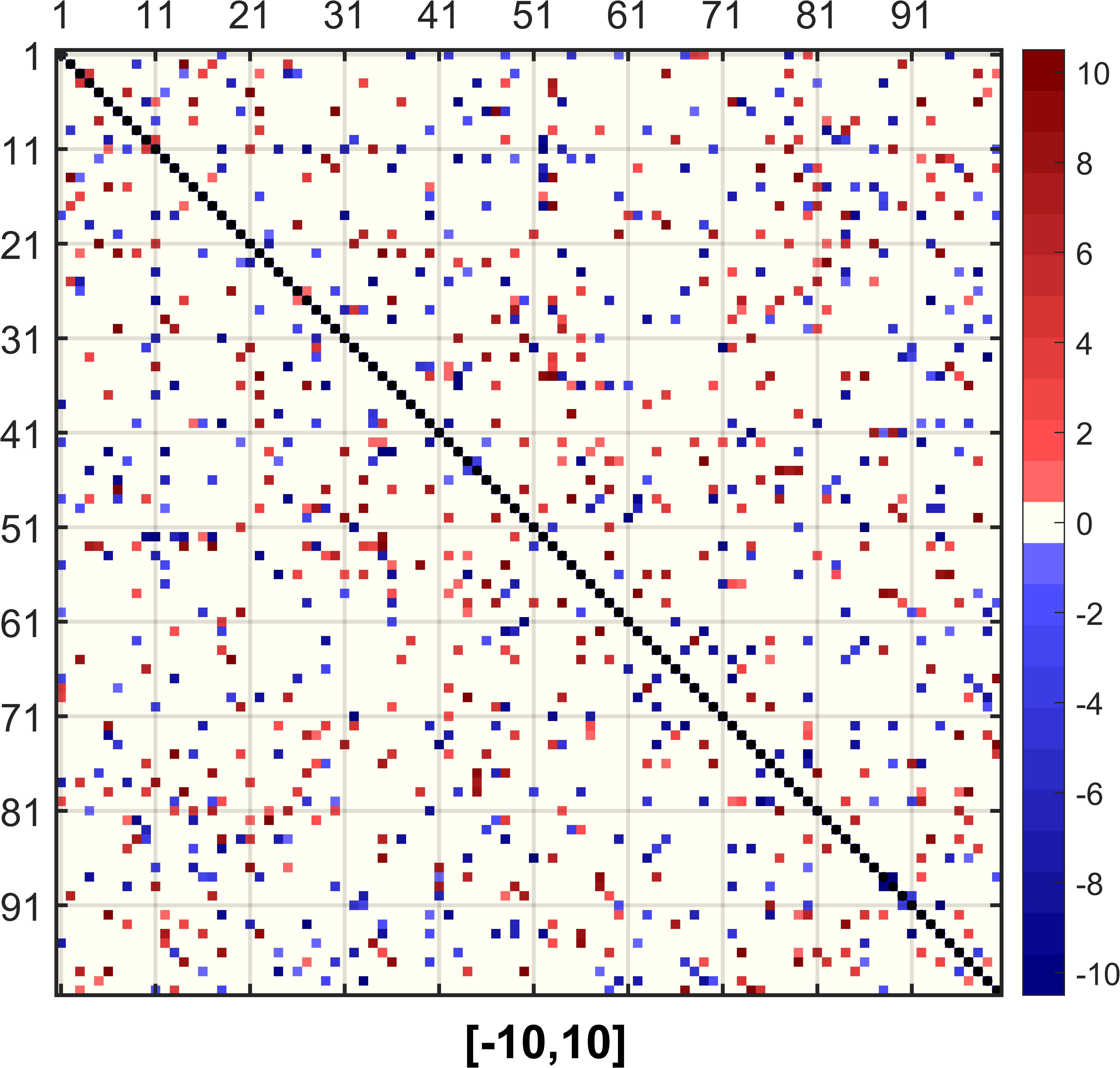}
        \caption{w01\_100.5}
    \end{subfigure}

    \vspace{0.5cm} 
    \begin{subfigure}[t]{0.55\textwidth}
        \centering
        \includegraphics[width=\linewidth]{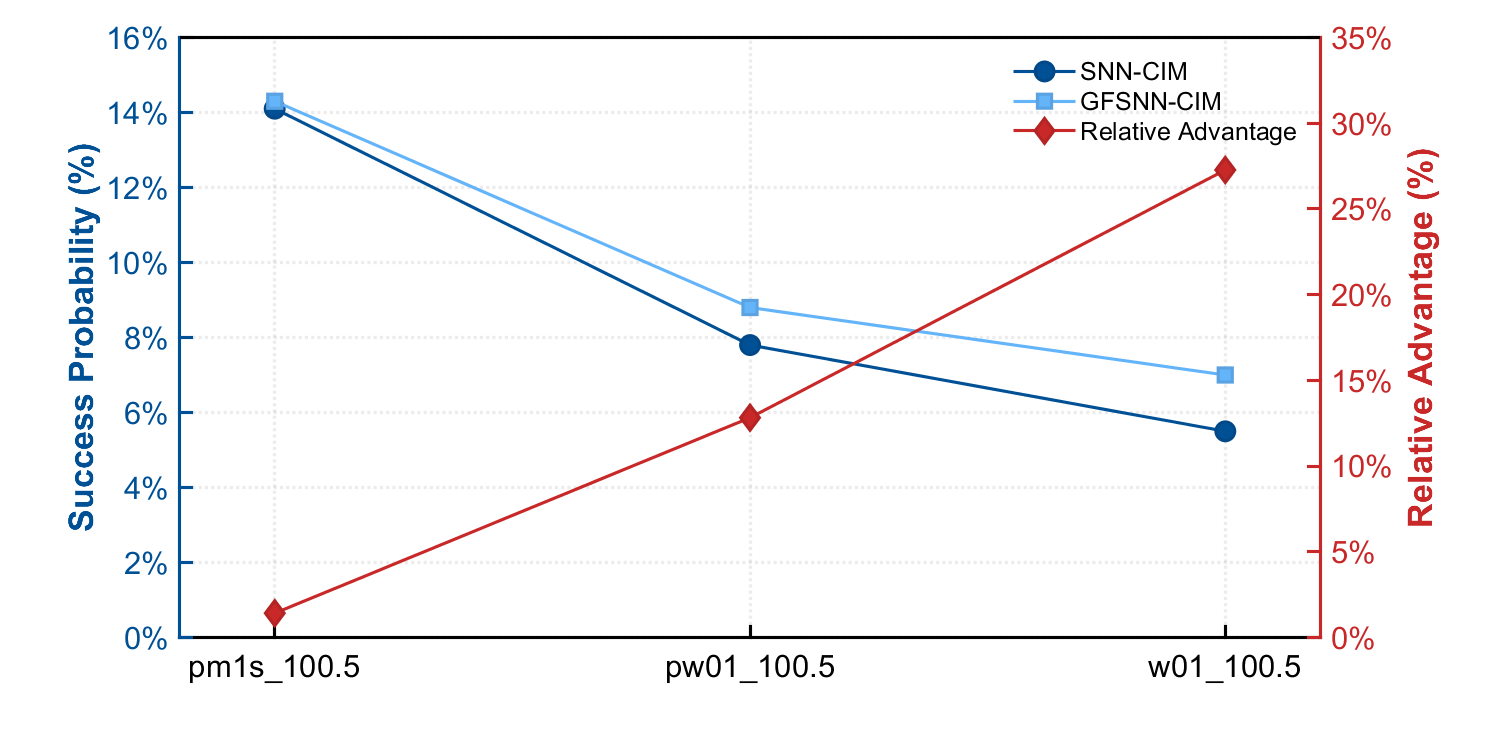}
        \caption{Performance comparison}
    \end{subfigure}

    \caption{
        (a) The weight distribution of pm1s\_100.5. 
        (b) The weight distribution of pw01\_100.5. 
        (c) The weight distribution of w01\_100.5. 
        (d) Performance comparison between GFSNN-CIM and SNN-CIM.
    }
\end{figure}

To systematically evaluate the performance advantages of GFSNN-CIM over SNN-CIM in complex coupling scenarios, we conduct numerical simulations on three standard max-cut problems (pm1s\_100.5, pw01\_100.5, w01\_100.5) \cite{wiegele2007biq}. These benchmark problems share identical network scale (100 nodes) and connection density (10\%), but exhibit distinct weight distribution characteristics. They feature weight intervals of [-1,1], [0,10], and [-10,10] respectively, following integer value distributions. To visually demonstrate the progressive complexity, Figure 1 (a)-(c) presents heat maps with unified color scales to display their coupling properties. In the figures, the zero-value diagonal (representing self-connections) is shown in black, other zero-value couplings in light yellow, with off-diagonal elements exhibiting symmetric distributions.

We compare the performance between GFSNN-CIM and SNN-CIM through numerical experiments, using the exact Max-Cut solutions provided by the datasets as evaluation criteria. Statistical analysis based on 1000 independent repeated trials (Figure 1 (d)) demonstrates that GFSNN-CIM exhibits performance advantages. In terms of ground-state convergence probability, GFSNN-CIM achieves up to 27\% improvement over SNN-CIM, with this advantage amplifying as problem complexity increases. While both algorithms show declining success rates with increasing problem complexity, GFSNN-CIM exhibits a slower degradation rate, indicating superior robustness in handling complex coupling problems.

\subsection{The results of TAP in the 5\(\times\)5 grid network}

The 5\(\times\)5 grid network is consists of 25 nodes and 80 links (as shown in Figure 2(a)), all links have identical attributes: freeFlow (\(t_a^0=1\)) and traffic capacity (\(f_{a,{\text{capacity}}}=25\)). This grid network has an initial traffic flow distribution (Figure 2(b)). Fifteen OD pairs are defined with the traffic demand for each OD pair $D_p=5$. Specifically, these are: (1,25,5), (2,20,5), (3,23,5), (4,24,5), (5,21,5), (6,24,5), (10,16,5), (11,15,5), (15,11,5), (16,10,5), (20,2,5), (21,5,5), (23,3,5), (24,6,5), (25,1,5). Here, the first, second, and third numbers in brackets represent the origin, destination, and the traffic demands $D_p$ for each OD pair, respectively.

\begin{figure}[htbp!]
    \centering
    \begin{subfigure}[t]{0.32\textwidth}
        \centering
        \includegraphics[width=\linewidth]{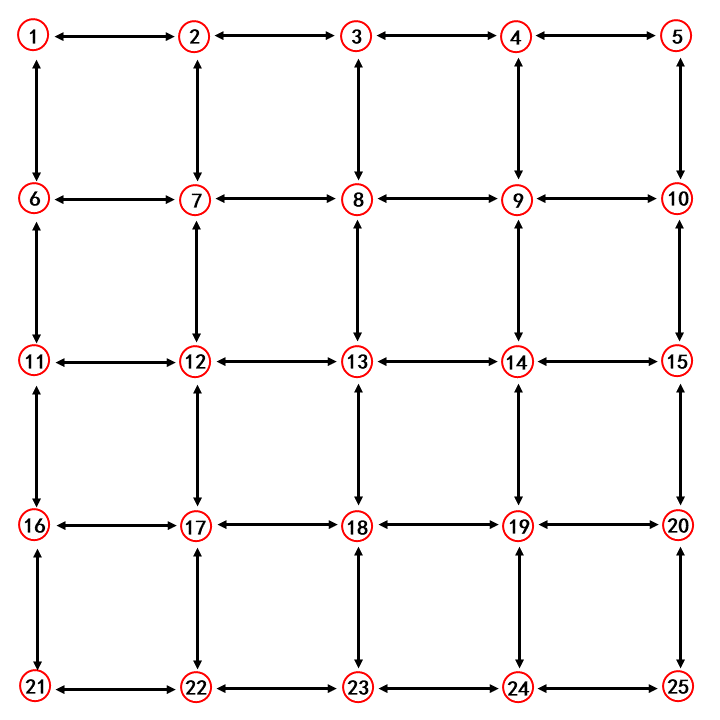}
        \caption{The grid network.}
    \end{subfigure}
    \hspace{0.05\textwidth}
    \begin{subfigure}[t]{0.41\textwidth}
        \centering
        \includegraphics[width=\linewidth]{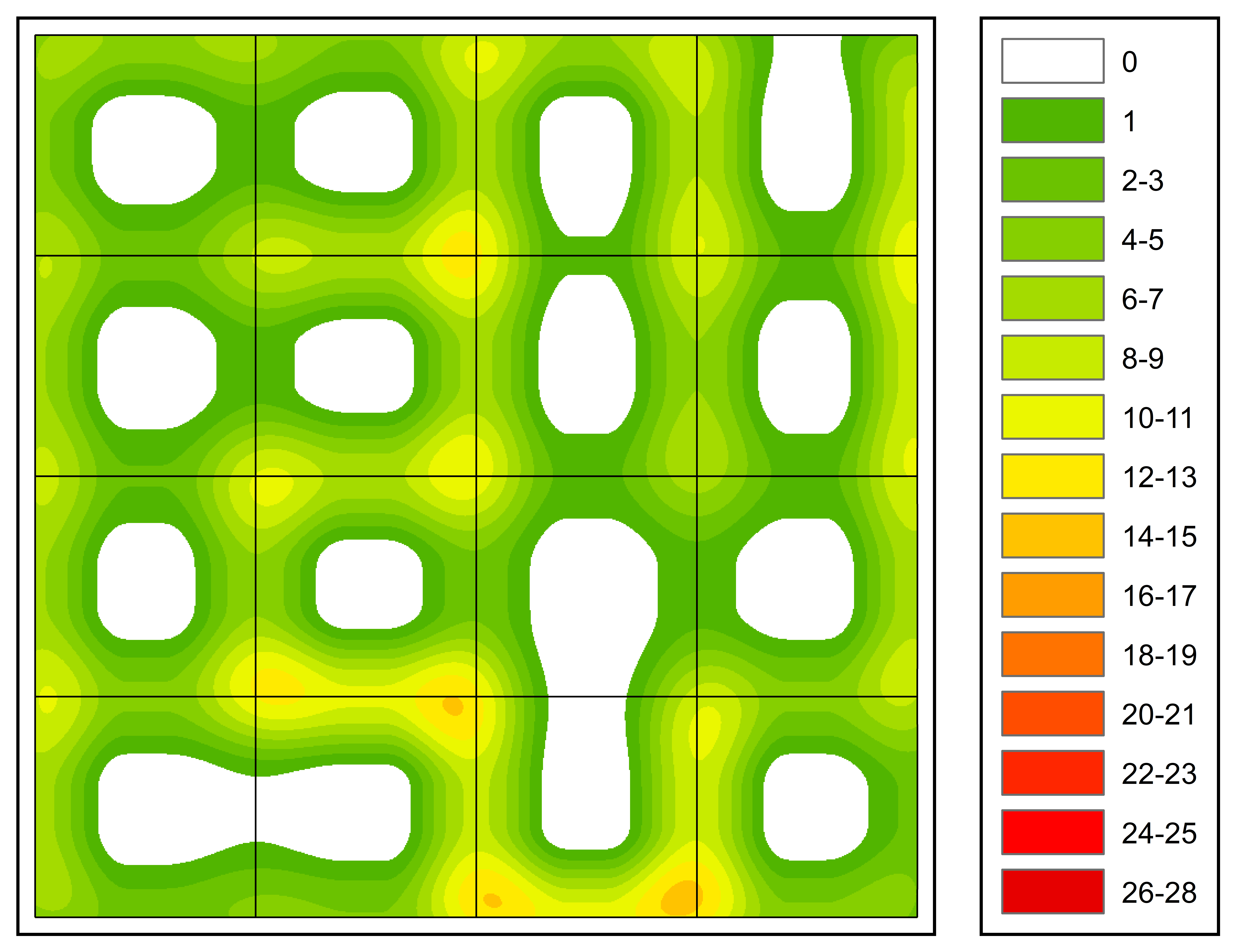}
        \caption{Initial flow.}
    \end{subfigure}
    \caption{(a) The $5\times 5$ grid network. (b) Heat map of initial flow in the 5\(\times\)5 grid network.}
\end{figure}
   
    We design two comparative experiment to evaluate algorithm performance. The experiment employs 1-vehicle discretization units (75 vehicle groups with 1 vehicle each), solved respectively by GFSNN-CIM, SNN-CIM, and simulated annealing (SA), corresponding to an Ising model with 226 spins. The second experiment utilizes finer 0.1-vehicle units (750 groups with 0.1 vehicle each), computed via GFSNN-CIM solver, scaling the Ising model to 2,251 spins. In both experiments, all vehicle groups within the same OD pair are assigned identical three alternative routes \{$R_1$, $R_2$, $R_3$\} ($R_1$: DIA-recommended routes; $R_2$: the shortest-distance route under free-flow conditions; $R_3$: a detour route.) The high-approximation quadratic function used has \(\gamma_{a,1}=0.002708\), \(\gamma_{a,2}=0.960831\), and \(\gamma_{a,3}=0.036461\) (its relative difference from the objective function is \(\pm 0.5\%\)). Additionally, the FW algorithm and DIA algorithm are incorporated to provide reference solutions. The FW algorithm provides solutions under continuous flow conditions, where 500 iterations are performed to ensure full convergence. The DIA algorithm employs a sequential assignment strategy, conducting 1000 simulation trials with different assignment orders to select the optimal solution. 

\begin{figure}[htbp!]
    \centering
    \begin{subfigure}[t]{0.295\textwidth}
        \centering
        \includegraphics[width=\linewidth]{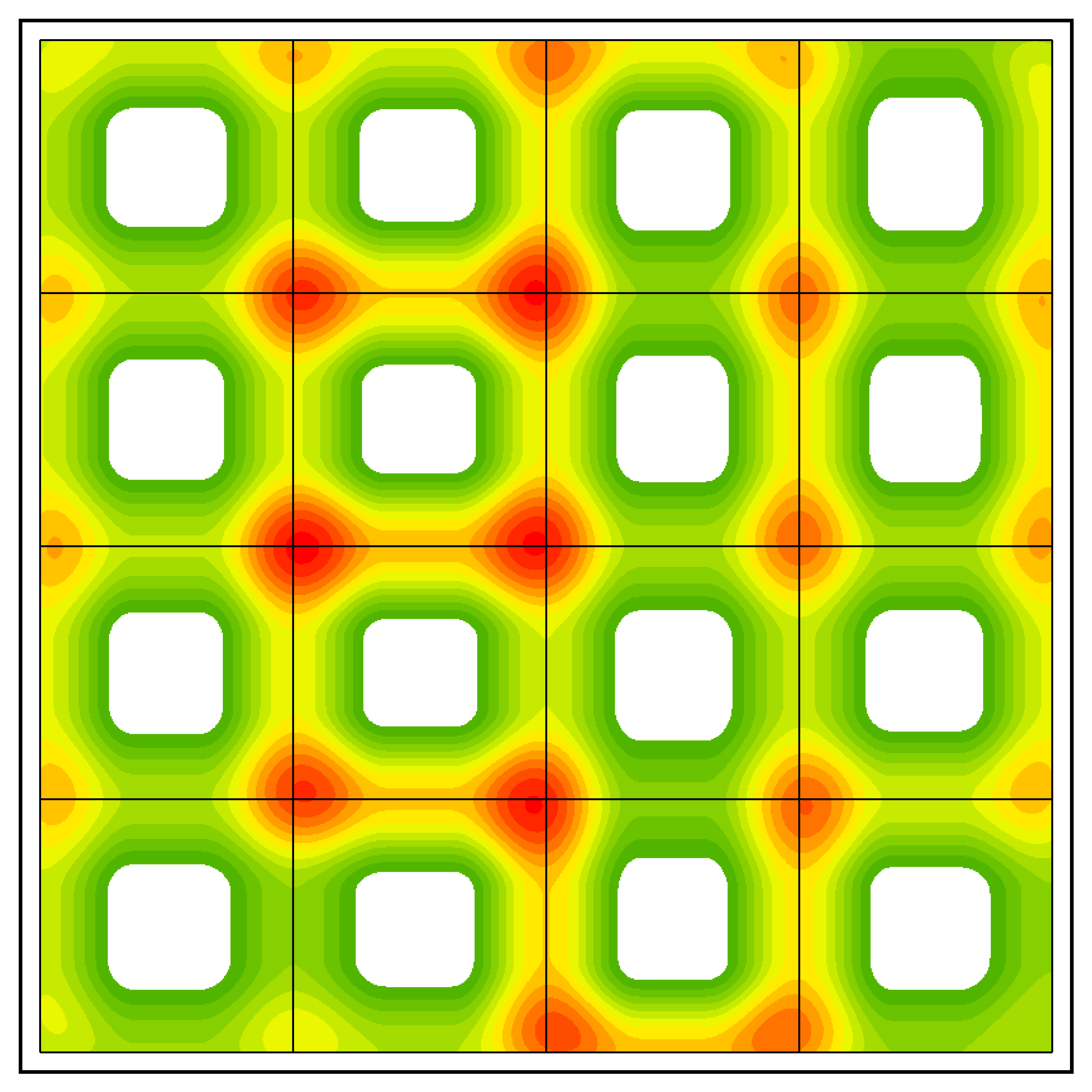}
        \caption{FW (continuous)}
    \end{subfigure}
    \hfill
    \begin{subfigure}[t]{0.295\textwidth}
        \centering
        \includegraphics[width=\linewidth]{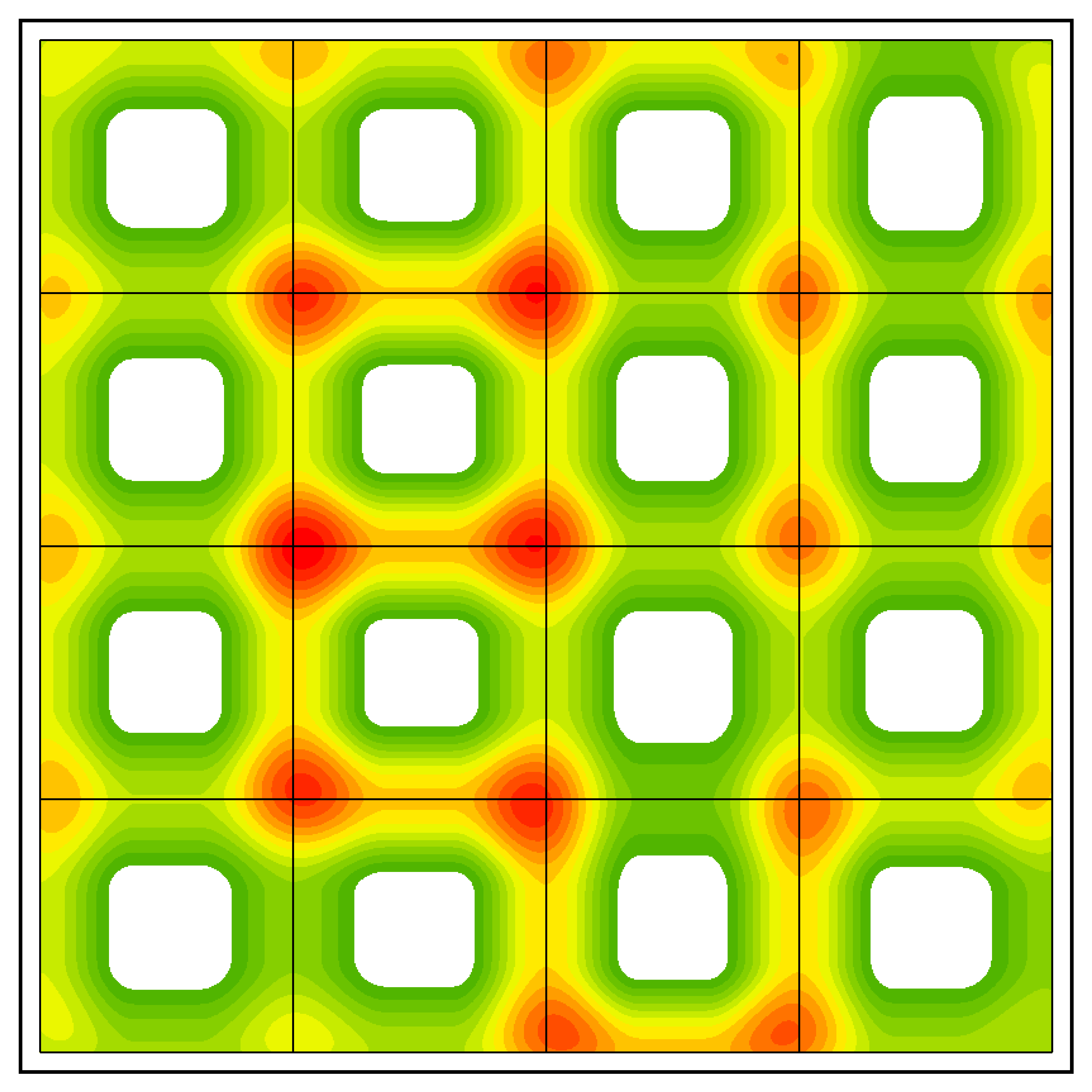}
        \caption{DIA (1-vehicle)}
    \end{subfigure}
    \hfill
    \begin{subfigure}[t]{0.388\textwidth}
        \centering
        \includegraphics[width=\linewidth]{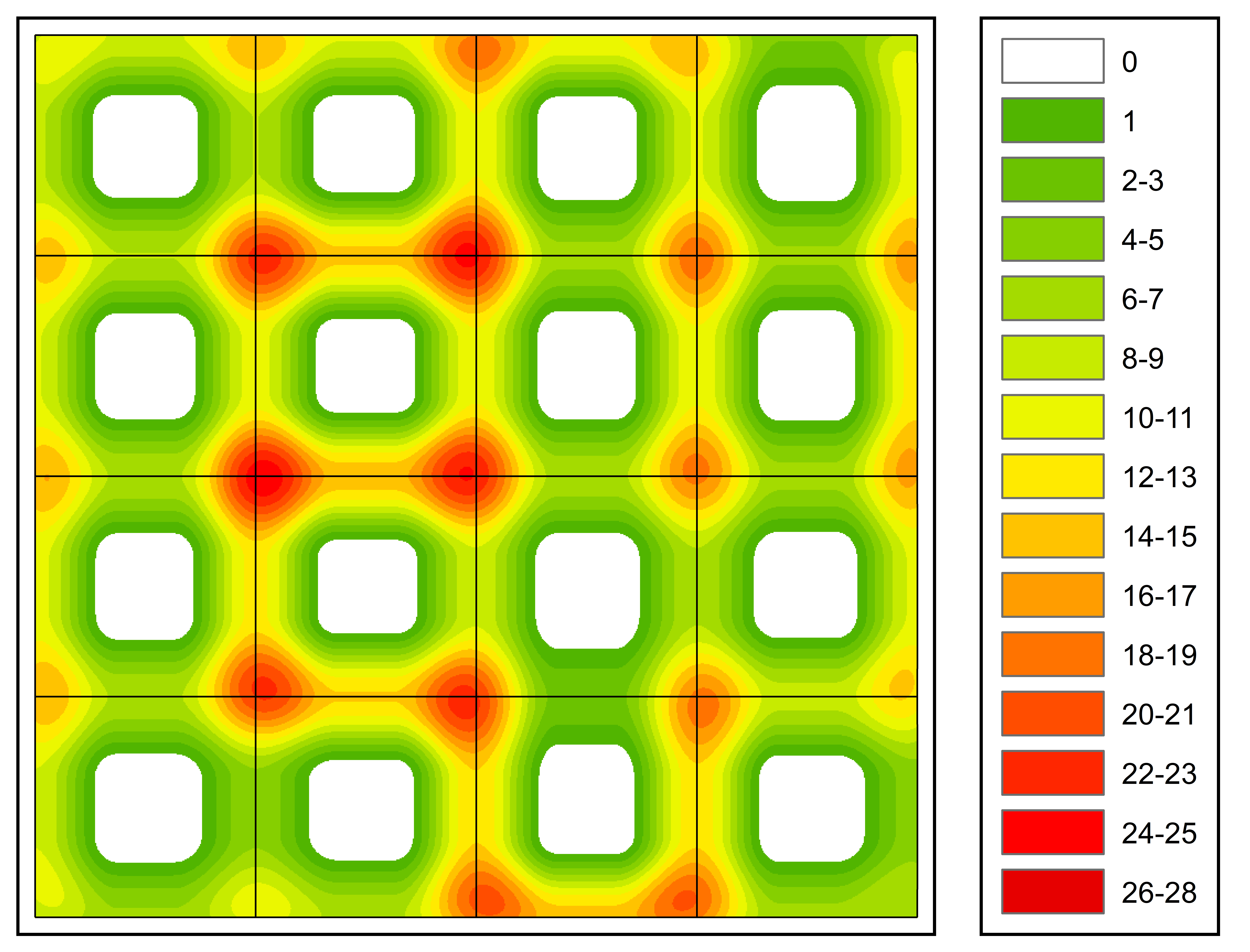}
        \caption{GFSNN-CIM (1-vehicle)}
    \end{subfigure}
    \hfill
    
    \centering
    \begin{subfigure}[t]{0.295\textwidth}
        \centering
        \includegraphics[width=\linewidth]{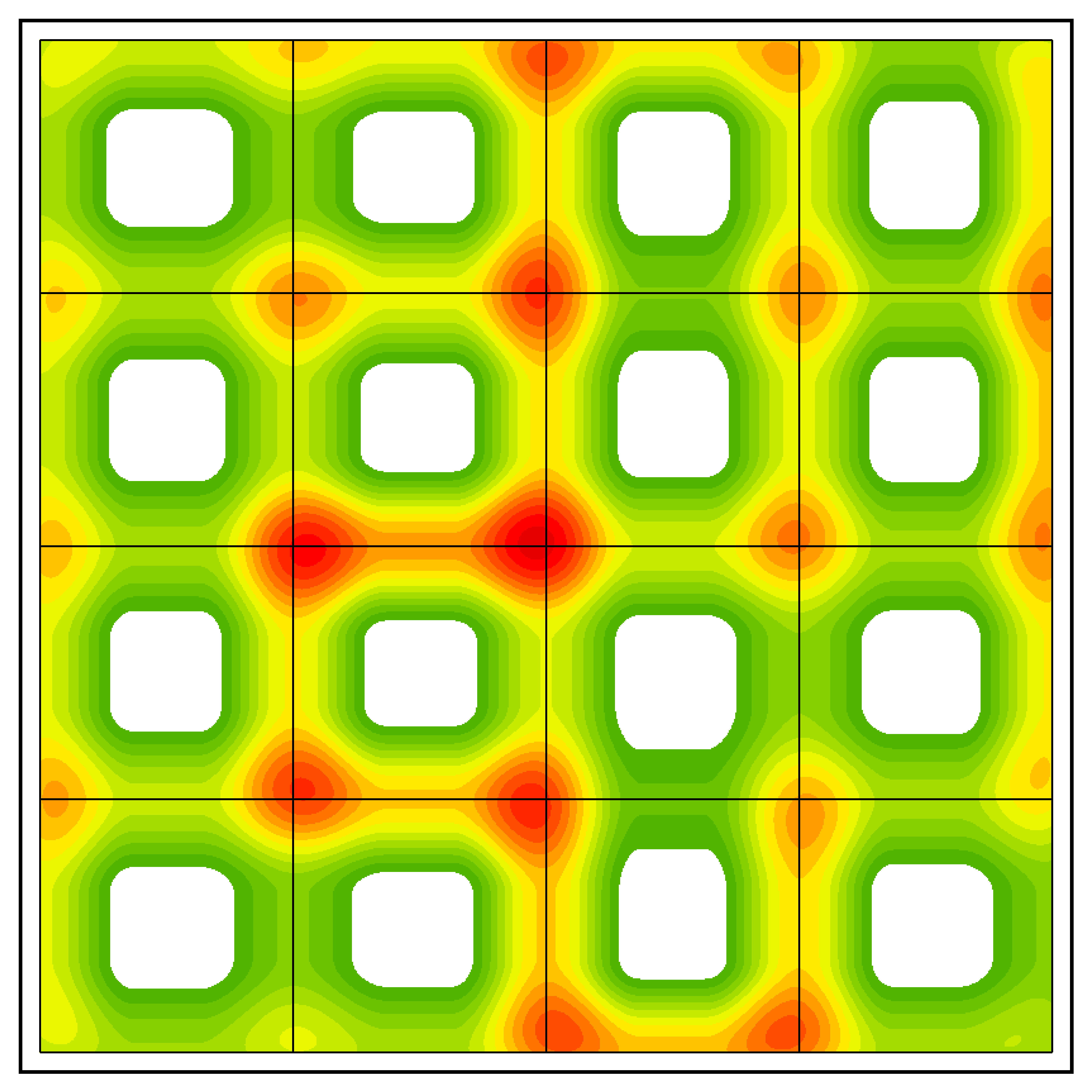}
        \caption{SA (1-vehicle)}
    \end{subfigure}
    \hfill
    \begin{subfigure}[t]{0.295\textwidth}
        \centering
        \includegraphics[width=\linewidth]{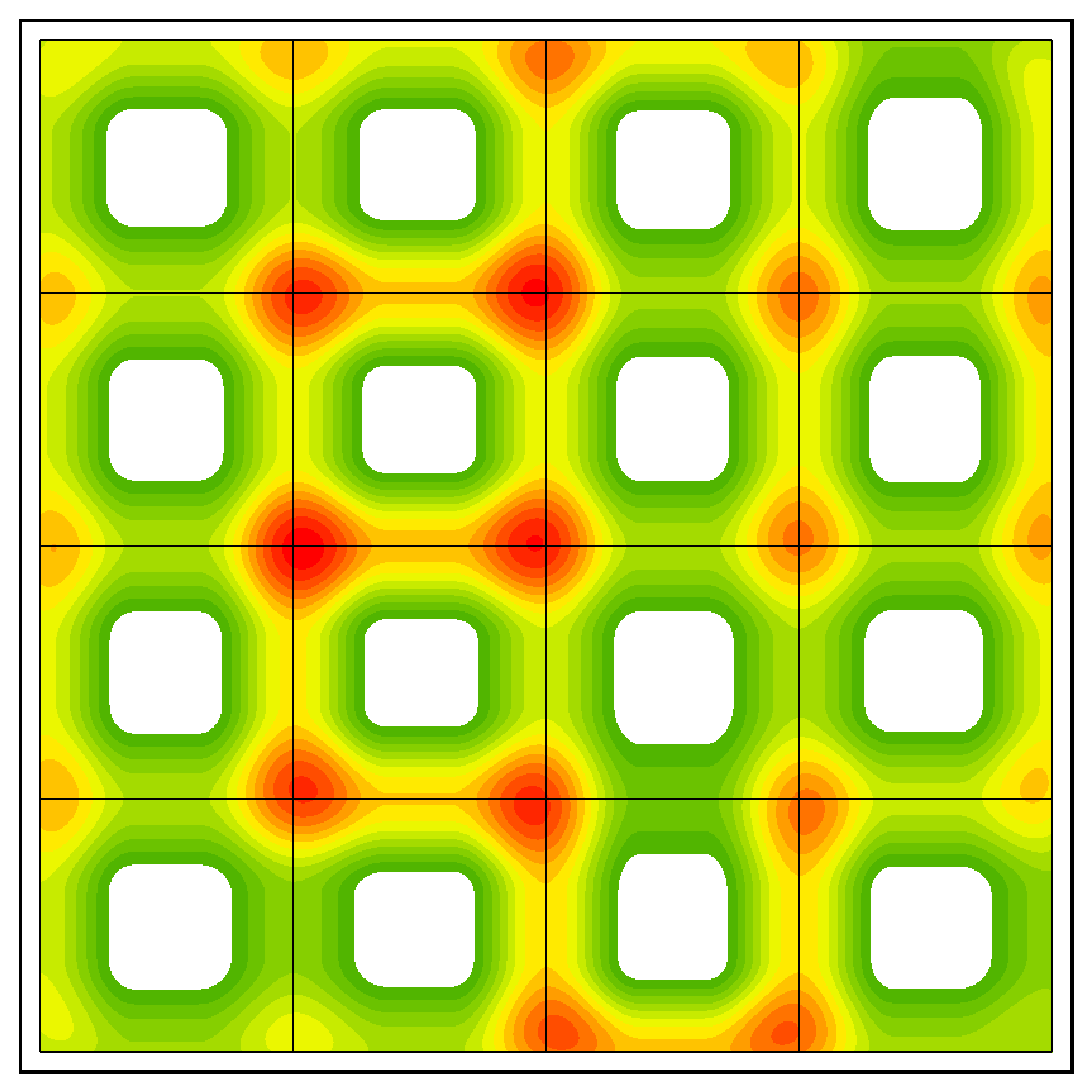}
        \caption{SNN-CIM (1-vehicle)}
    \end{subfigure}
      \hfill
    \begin{subfigure}[t]{0.388\textwidth}
        \centering
        \includegraphics[width=\linewidth]{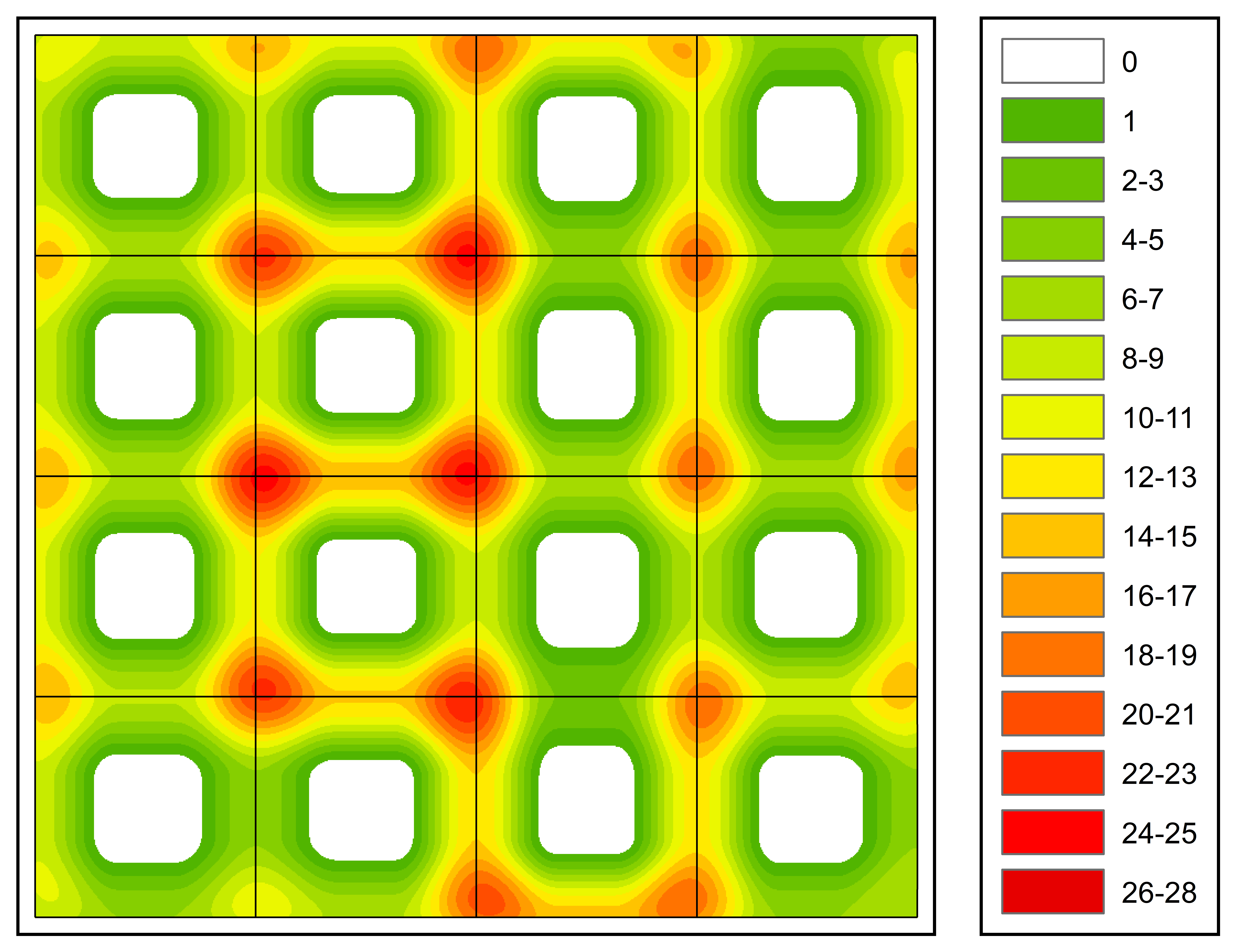}
        \caption{GFSNN-CIM (0.1-vehicle)}
    \end{subfigure}
\caption{(a) The heat map of FW solution. (b) The heat map of DIA solution. (c) The heat map of GFSNN-CIM solution (1-vehicle units). (d) The heat map of SA solution (1-vehicle units). (e) The heat map of SNN-CIM solution (1-vehicle units). (f) The heat map of GFSNN-CIM solution (0.1-vehicle units).}
\end{figure}

Traffic flow distribution is visualized using heat maps with a green-yellow-red color gradient, where green indicates smooth conditions and red represents traffic congestion. As shown in Figure 2(b), the initial flow distribution in the 5×5 grid network exhibits uniform green tones, demonstrating smooth operation across the network. Comparative analysis of the experimental results in Figure. 3 reveals that all algorithms significantly increase red areas compared to the initial state, confirming effective traffic demand assignment. Notably, the SA algorithm results (Figure 3(d)) display distinct anomalies, with a dark red cluster forming at the central node (5,5). This phenomenon visually demonstrates the SA algorithm's performance deficiency in balanced flow assignment.

The distinct route selection strategies among algorithms can be illustrated through the third vehicle group under traffic demand (1,25,5). This vehicle group's alternative route set was \{$R_1^{1-25-3}$, $R_2^{1-25-3}$, $R_3^{1-25-3}$\}. These routes can be represented by the node links in Figure 2 (a):\\

$R_1^{1-25-3}$: 1$\rightarrow$2$\rightarrow$7$\rightarrow$8$\rightarrow$9$\rightarrow$14$\rightarrow$19$\rightarrow$20$\rightarrow$25;\\

$R_2^{1-25-3}$: 1$\rightarrow$2$\rightarrow$3$\rightarrow$4$\rightarrow$5$\rightarrow$10$\rightarrow$15$\rightarrow$20$\rightarrow$25;\\

$R_3^{1-25-3}$: 1$\rightarrow$6$\rightarrow$11$\rightarrow$16$\rightarrow$21$\rightarrow$22$\rightarrow$23$\rightarrow$24$\rightarrow$25.\\

Results demonstrate that DIA, SNN-CIM, and GFSNN-CIM algorithms all select $R_1^{1-25-3}$, whereas the SA algorithm chooses the suboptimal $R_3^{1-25-3}$. This indicates that the SA algorithm suffers from local optimum entrapment, while GFSNN-CIM avoids this pitfall through its global mean-amplitude feedback mechanism. 

\begin{table}[htbp!]
\centering
\caption{The performance comparison of TAP in the 5\(\times\)5 grid network.}
\label{tab:baseline}
\setlength{\tabcolsep}{7.5pt}
\begin{tabular}{@{}llll@{}}
\toprule
\textbf{Flow Type} & \textbf{Algorithm} & \textbf{Objective function}& \textbf{$\Delta$ vs Reference (\%)} \\
\midrule
Continuous & FW & 1308.33 & Reference \\ 
\addlinespace
\multirow{3}{*}{Discrete (1-vehicle)}& DIA & 1308.85 & 0.0397 (vs FW)\\
& SA & 1311.79 & 0.2645 (vs FW)\\
& SNN-CIM& 1308.78 & 0.0344 (vs FW)\\
& GFSNN-CIM& 1308.78 &0.0344 (vs FW)\\
\addlinespace
\multirow{3}{*}{Discrete (0.1-vehicle)}
   & FW'& 1308.35& Reference \\
   & SNN-CIM& 1308.55 & 0.0153 (vs FW')\\
   & GFSNN-CIM& 1308.48& 0.0099 (vs FW')\\
\bottomrule
\end{tabular}
\footnotesize
\begin{flushleft}
*$\Delta=(\text{Method}-\text{Reference})/\text{Reference}\times100\%$.\\
*FW': Rounding continuous FW results to one decimal place.\\
\end{flushleft}
\vspace{2mm}
\end{table}

Table 1 presents the algorithmic performance comparison for STP in the 5×5 grid network. The UE solution obtained by the FW algorithm under continuous flow conditions serves as the benchmark. With 1-vehicle discretization units, GFSNN-CIM achieves superior solution accuracy, exhibiting merely 0.0344\% deviation from the FW result lower than the deviations of DIA (0.0397\%) and SA (0.2645\%). When adopting finer 0.1-vehicle units, GFSNN-CIM further improves its precision, reducing the deviation to 0.0115\% against the FW benchmark and to 0.0099\% compared to the rounded FW' solution (1308.35).

\subsection{ The results of TAP in the Beijing's urban road network}

\begin{figure}[htbp]
    \centering
    \begin{subfigure}[t]{0.48\textwidth}
        \centering
        \includegraphics[width=\linewidth]{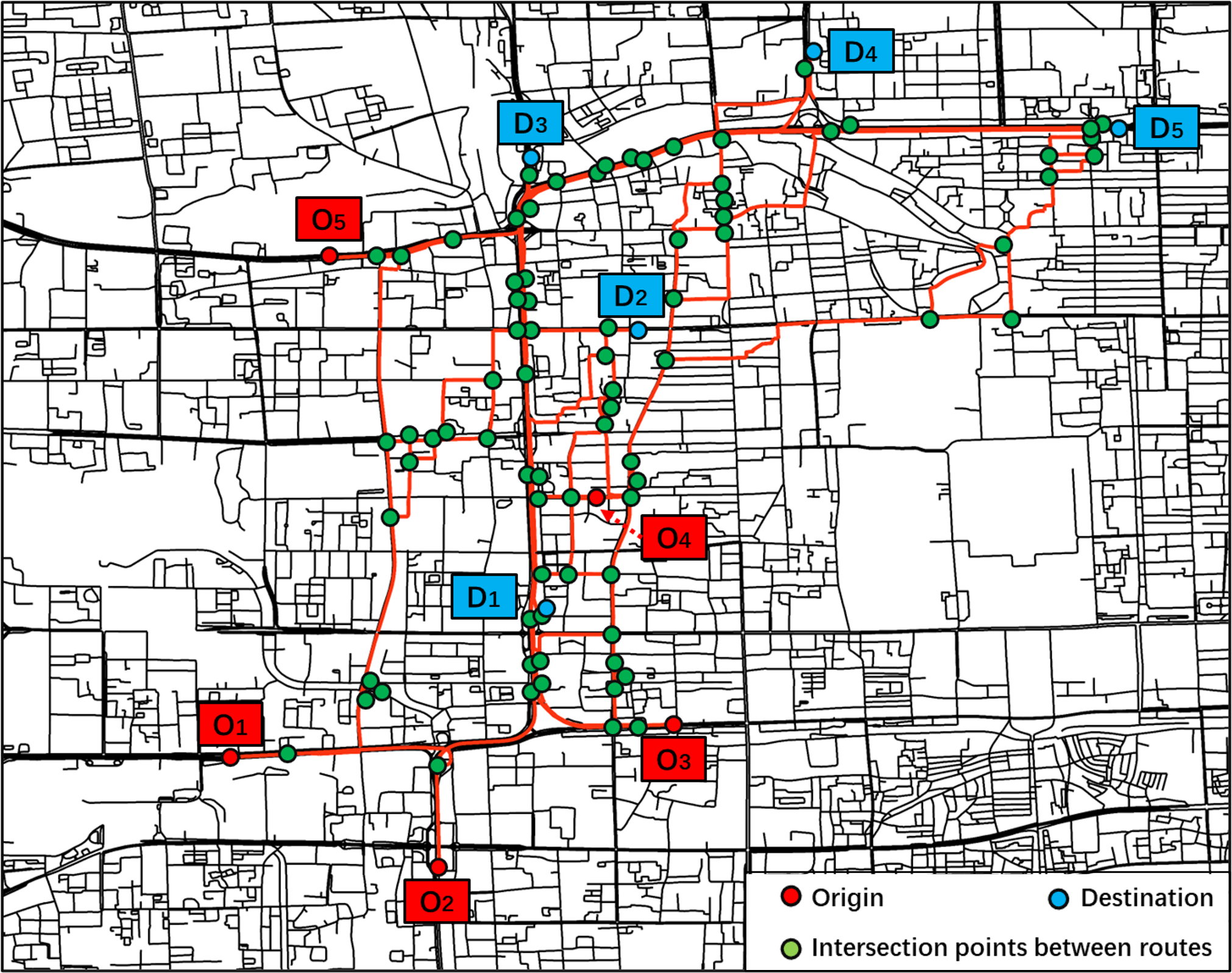}
        \caption{Part of the map of Beijing city}
    \end{subfigure}
    \hfill
    \begin{subfigure}[t]{0.499\textwidth}
        \centering
        \includegraphics[width=\linewidth]{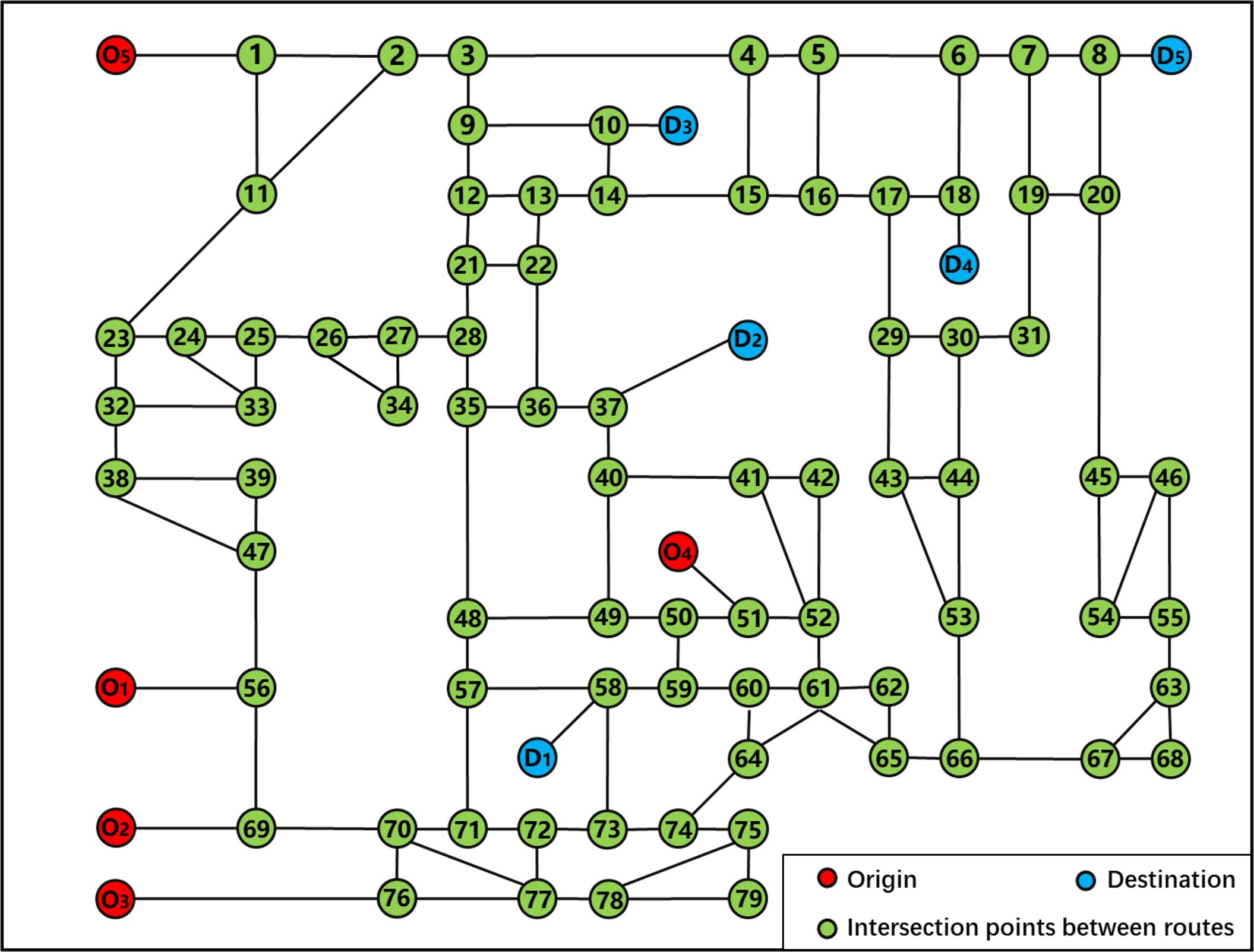}
        \caption{Equivalent network}
    \end{subfigure}
    \hfill
    \caption{(a) Part of the map within the Third Ring Road of Beijing. (b) Equivalent network of Beijing road network.}
\end{figure}
To validate the efficacy and generalizability of our research, we selected a real-world road network comprising 408 streets (links) within Beijing's Third Ring Road (highlighted by red lines in Figure 4(a)) for the TAP. Compared to the 5×5 grid network, this Beijing urban road network exhibits larger scale, more complex topology, and realistic traffic data characteristics. OD pairs are marked by red (origins) and blue (destinations) dots in Figure 4(a). For intuitive display, this road network was drawn into an equivalent network with 89 nodes (Figure 4 (b)), where the green nodes are the intersections between alternative vehicles' routes, and each section between two nodes consists of multiple links. The capacity of each link was set to 1600, and the freeFlow time was calculated by assuming a speed of 16.67 m/s (60 km/h).

Figure 5(a) displays the Beijing evening rush-hour traffic flow heat map derived from the T-drive data set \cite{yuan2011t}. The initial flow condition (Figure 5(b)) was established by removing 1600 vehicles between BeijingXi Railway Station ($O_1$) and Gulou Street ($D_5$) from the T-drive data set. We defined 20 OD pairs with a total traffic demand of 1600, specified as follows: (1,2,30), (1,3,30), (1,4,10), (1,5,30), (2,1,140), (2,2,60), (2,3,80), (2,4,30), (2,5,50), (3,1,120), (3,2,40), (3,3,60), (3,4,30), (3,5,10), (4,2,120), (4,3,120), (4,4,80), (4,5,80), (5,4,180), (5,5,300). These demands were divided into $N=160$ vehicle groups, with each group representing $g=10$ vehicles. This problem was solved by GFSNN-CIM, SNN-CIM, and SA, corresponding to an Ising model with 481 spins. 

In this problem, the parameters on different links are different, which means that multiple quadratic functions are used. Its solving process consists of two steps: (1) a common approximate quadratic function (\(\gamma_{a,1}=0.00001484, \gamma_{a,2}=0.99167356x\) and \(\gamma_{a,3}=0.00831161\), with a relative error of 0.002\%) is used for all links,  and GFSNN-CIM provides the predicted flow (GFSNN-CIM1). (2) Based on the predicted flows, the more accurate function parameters are determined for each link (the relative errors are at the 10\textsuperscript{-10} level). The optimal solution can be obtained after 1000 calculations using GFSNN-CIM (GFSNN-CIM2). This two-step approach balances computational efficiency and accuracy. SNN-CIM and SA solves \(Q^{\prime~T}JQ^{\prime}\) in the same way as the second step of GFSNN-CIM and also runs 1000 times to find the optimal solution. The DIA algorithm also follows the sequential allocation approach in the previous section and selects the best result from 1,000 calculations.

Figures 5(c)-(f) present the flow distributions obtained using the DIA, SA, SNN-CIM, and GFSNN-CIM2. Compared to Figure 5(a), the effectiveness of these algorithms is clearly demonstrated by the significant mitigation of traffic congestion within the blue box region. In particular, flow redistribution toward the red box region induces a chromatic transition from colorless to green, indicating an increase in traffic density.

\begin{figure}[htbp]
    \centering
    \begin{subfigure}[t]{0.32\textwidth}
        \centering
        \includegraphics[width=\linewidth]{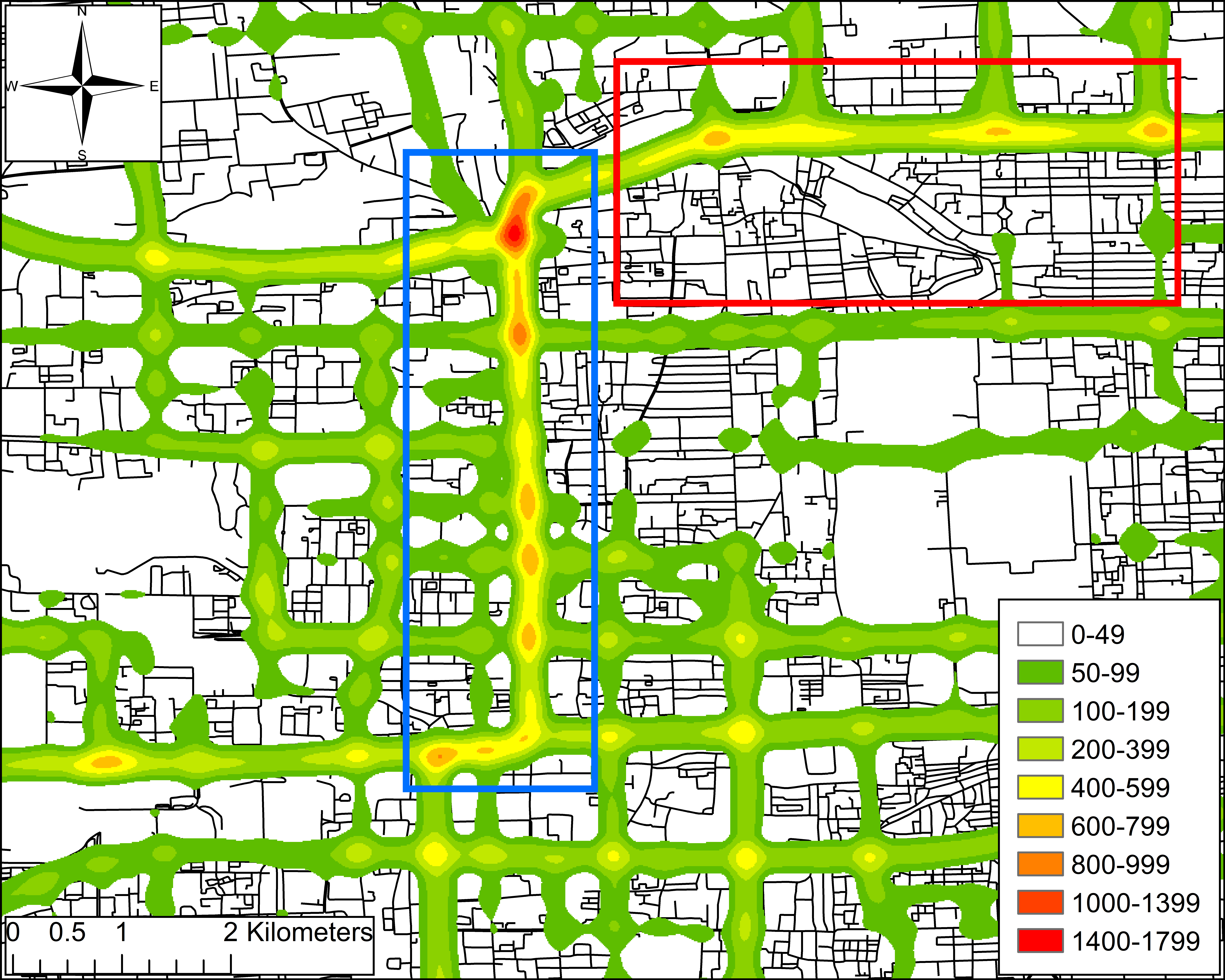}
        \caption{Beijing evening rush-hour heat map}
    \end{subfigure}
    \hfill
    \begin{subfigure}[t]{0.32\textwidth}
        \centering
        \includegraphics[width=\linewidth]{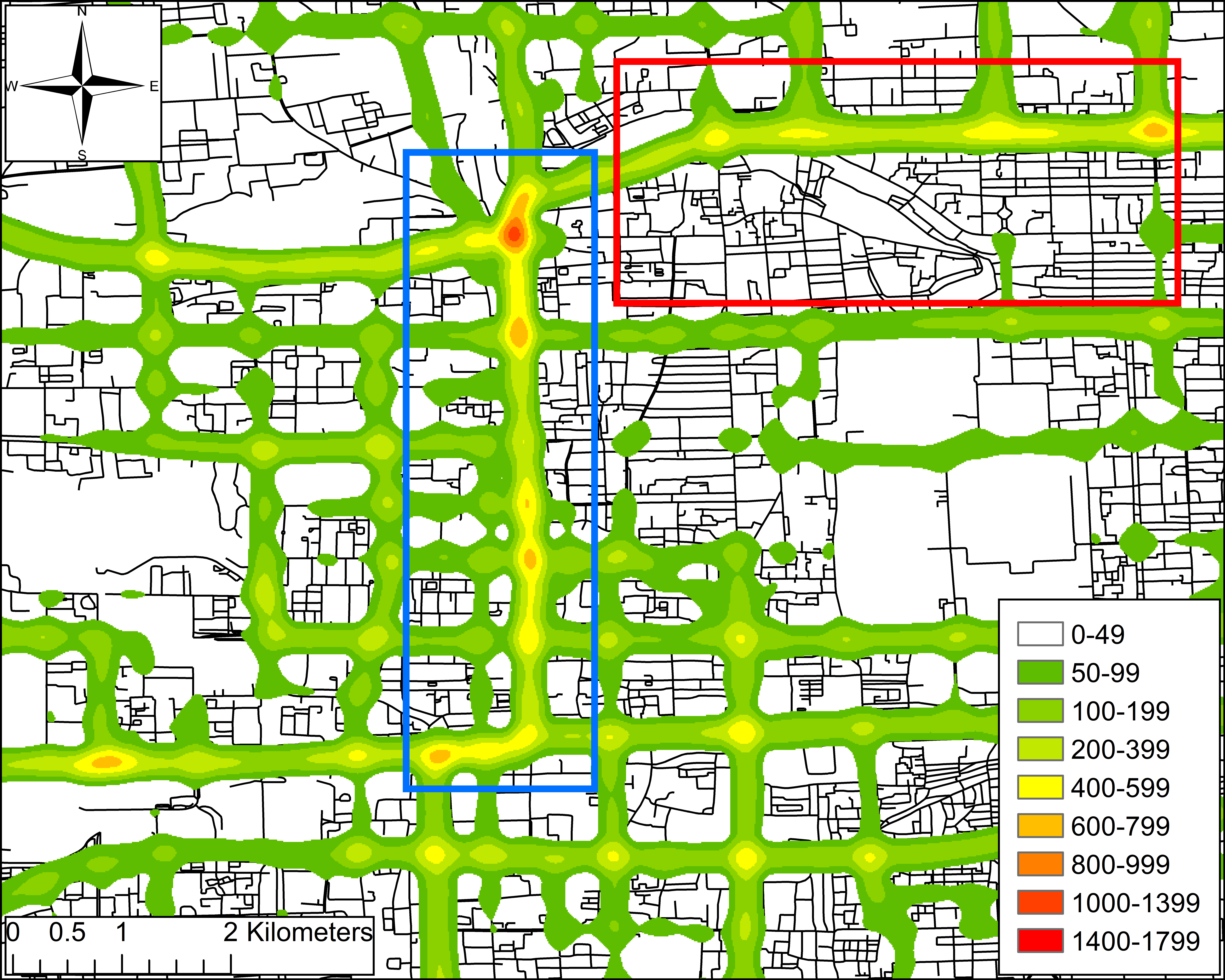}
        \caption{Initial flow condition}
    \end{subfigure}
    \hfill
    \begin{subfigure}[t]{0.32\textwidth}
        \centering
        \includegraphics[width=\linewidth]{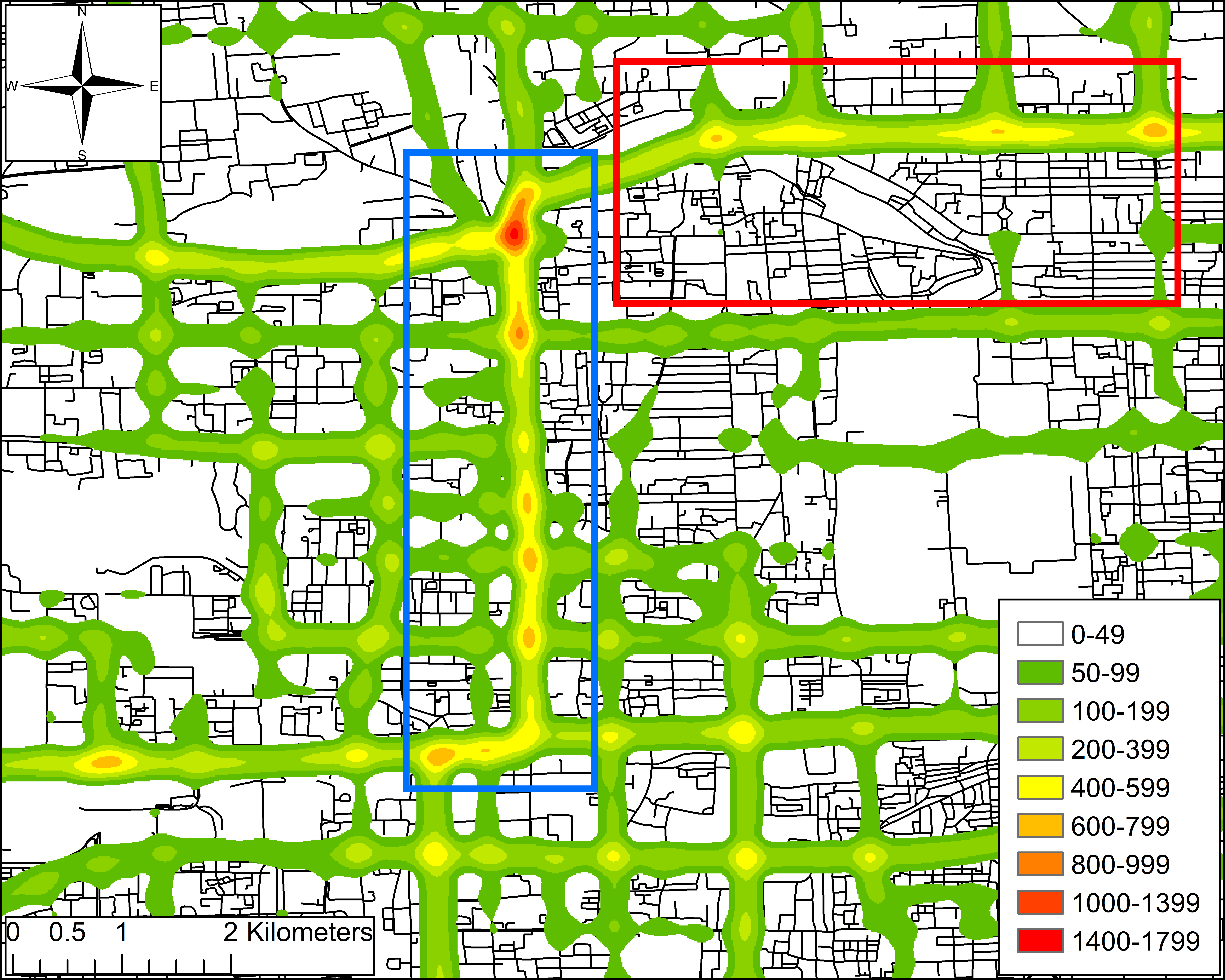}
        \caption{DIA}
    \end{subfigure}
    \hfill
    \centering
    \begin{subfigure}[t]{0.32\textwidth}
        \centering
        \includegraphics[width=\linewidth]{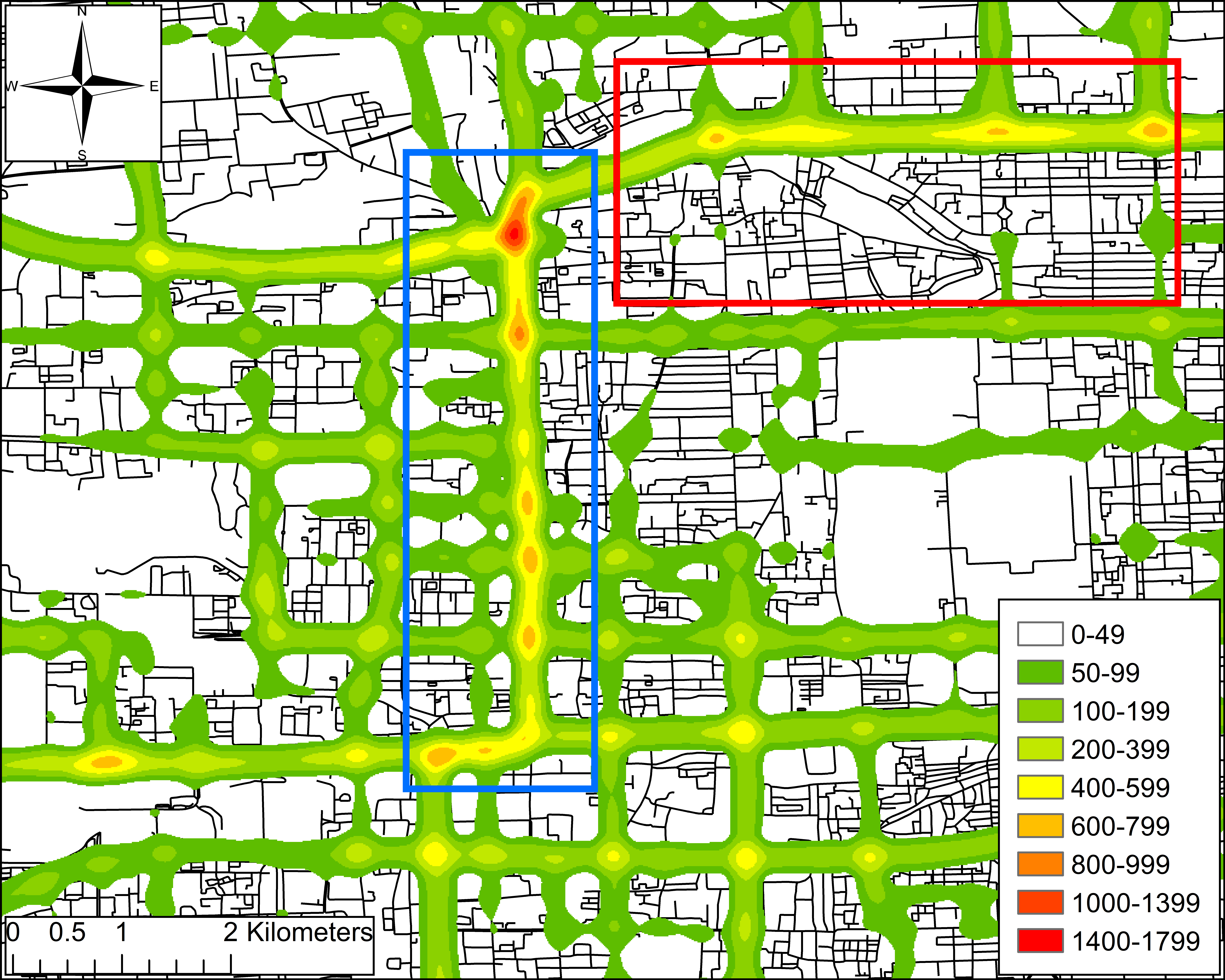}
        \caption{SA}
    \end{subfigure}
    \hfill
    \begin{subfigure}[t]{0.32\textwidth}
        \centering
        \includegraphics[width=\linewidth]{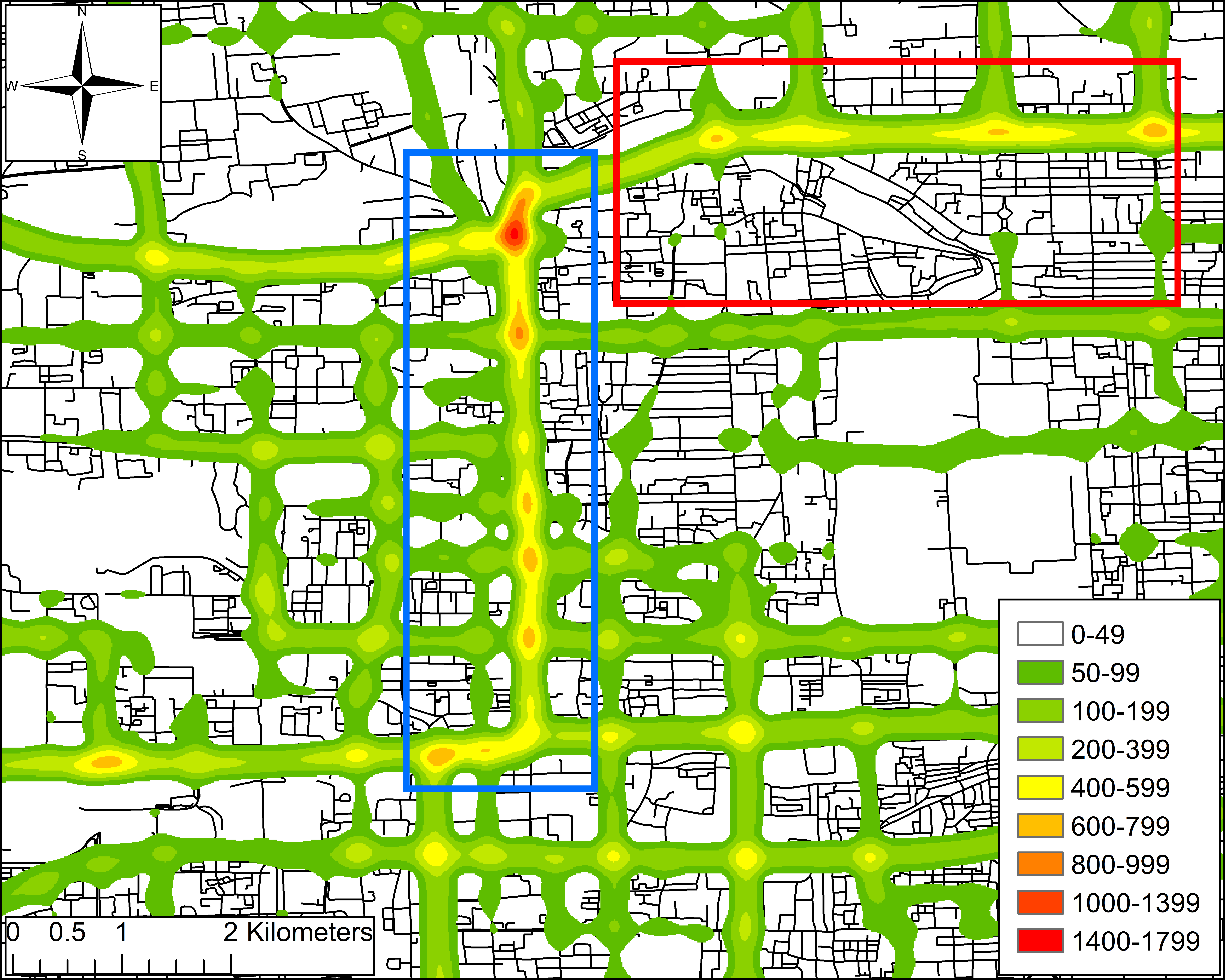}
        \caption{SNN-CIM}
    \end{subfigure}
    \hfill
    \begin{subfigure}[t]{0.32\textwidth}
        \centering
        \includegraphics[width=\linewidth]{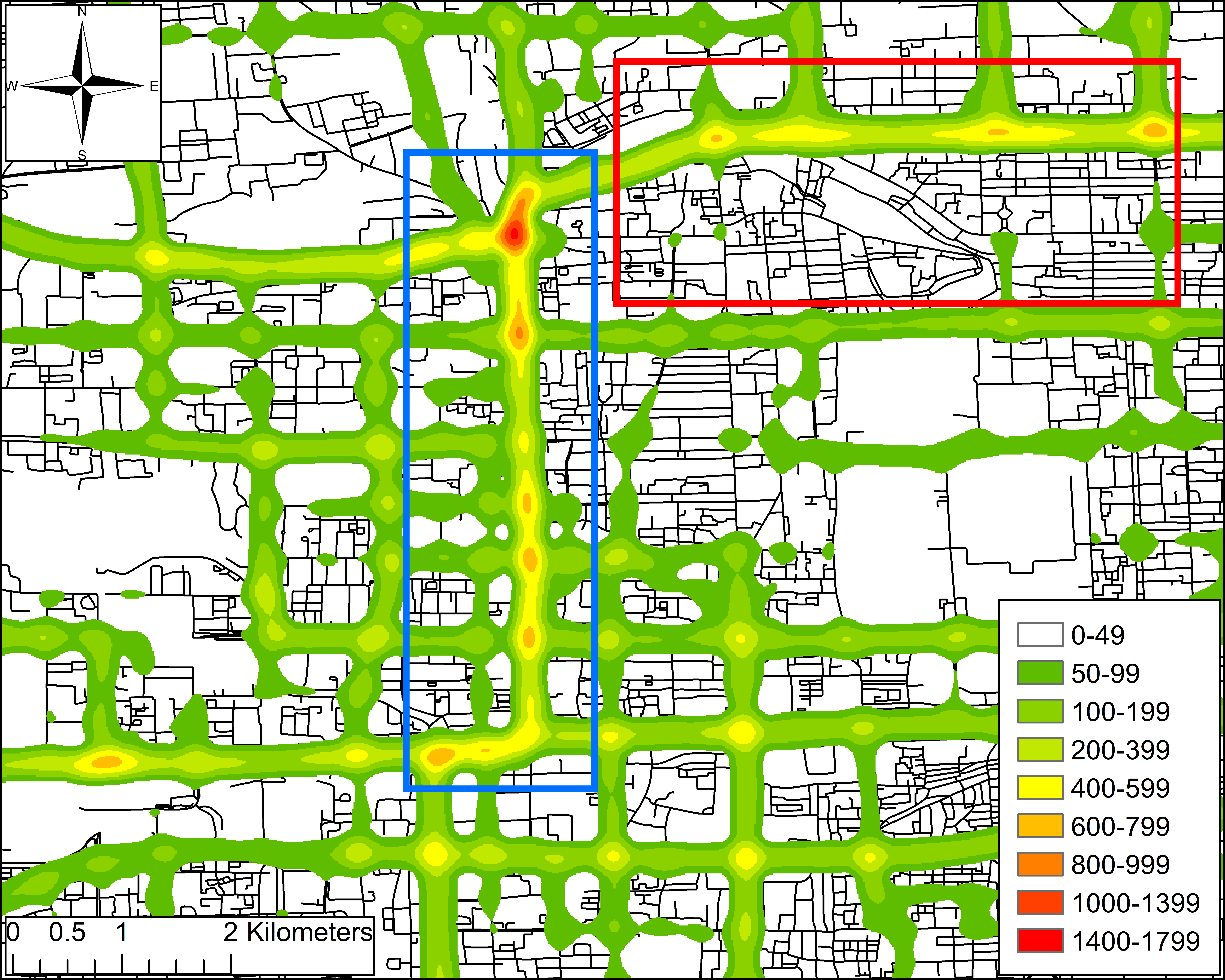}
        \caption{GFSNN-CIM2}
    \end{subfigure}
    \hfill
    \caption{(a) The heat map of Beijing traffic flow during the evening rush (16:00-20:59). (b) The Heat map of the initial flow condition. (c) The heat map of DIA solution. (d) The heat map of SA solution. (e) The heat map of SNN-CIM solution. (f) The heat map of GFSNN-CIM2 solution. }
\end{figure}

Table 2 compares the objective function values of DIA, SA, SNN-CIM, GFSNN-CIM1, and GFSNN-CIM2. GFSNN-CIM2 outperforms all comparative algorithms in optimization performance, demonstrating superior global optimization capability within constrained route sets. A case study of the sixth vehicle group under traffic demand (3,1,120) provides deeper insights into algorithmic decision-making differences. The alternative route sets for this group is {$R_1^{3-1-6}$, $R_2^{3-1-6}$, $R_3^{3-1-6}$}, visualized as node-link diagrams in  Figure 4b ($R_1^{3-1-6}$:  the DIA-recommended route; $R_2^{3-1-6}$:  the shortest-distance route under free-flow conditions; $R_3^{3-1-6}$:  a detour route). \\

$R_1^{3-1-6}$: \(O_3\)$\rightarrow$76$\rightarrow$70$\rightarrow$77$\rightarrow$72$\rightarrow$73$\rightarrow$58$\rightarrow$\(D_1\).\\

$R_2^{3-1-6}$: \(O_3\)$\rightarrow$76$\rightarrow$70$\rightarrow$71$\rightarrow$57$\rightarrow$58$\rightarrow$\(D_1\).\\

$R_3^{3-1-6}$: \(O_3\)$\rightarrow$76$\rightarrow$77$\rightarrow$72$\rightarrow$73$\rightarrow$58$\rightarrow$\(D_1\). \\

The route selection strategies are: DIA, SNN-CIM and GFSNN-CIM1 all select $R_1^{3-1-6}$, SA chooses $R_2^{3-1-6}$, while GFSNN-CIM2 uniquely selects $R_3^{3-1-6}$. Although $R_3^{3-1-6}$ shows inferior performance when evaluated individually, its coordination with other groups' route selections collectively contributes to forming high-quality solutions. This phenomenon reveals GFSNN-CIM2's distinctive collective optimization mechanism. In conclusion, the results on Beijing's urban road network confirm that GFSNN-CIM adapts to large-scale, complex traffic scenarios, demonstrating significant potential for smart city transportation management.

\begin{table}[htbp!]
\centering
\caption{The performance comparison of TAP in Beijing's urban road network.}
\label{tab:algorithm_comparison}
\begin{tabular}{@{} l *{5}{S[table-format=8.2]} @{}}
\toprule
\multicolumn{1}{c}{\textbf{Algorithm}} & 
\multicolumn{1}{c}{\textbf{DIA}} & 
\multicolumn{1}{c}{\textbf{SA}} & 
\multicolumn{1}{c}{\textbf{SNN-CIM}} & 
\multicolumn{1}{c}{\textbf{GFSNN-CIM1}} & 
\multicolumn{1}{c}{\textbf{GFSNN-CIM2}} \\
\midrule
$Obj(Q_{ij})$ & 55937372.31 & 55946082.21 & 55936374.44 & 55937850.81 & 55935033.75 \\
\bottomrule
\end{tabular}
\par\vspace{0.5em}
\end{table}

Currently, an experimental CIM system on the 10\textsuperscript{2} spins scale solves COPs in 2.37 milliseconds \cite{wen2023optical}, and we believe that with CIM the TAP task can be solved in a similar millisecond time frame.  As the scale increases, the system can incorporate expanded sets of alternative routes, enabling precise traffic flow management in large-scale urban road networks. Our scheme provides a novel and efficient computational tool capable of quickly addressing complex traffic management challenges.

\section{Conclusion}
This study proposes a global mean-amplitude feedback-enhanced spiking neural network CIM (GFSNN-CIM) for solving UE-compliant TAP. By incorporating the BPR function to dynamically characterize flow-impedance relationships, the proposed method achieves high-fidelity modeling of real-world traffic scenarios. Experimental results on both Max-Cut problems and real-world TAP demonstrate that the proposed global mean-field feedback mechanism enhances the performance of conventional SNN-CIM. Given that quantum-inspired CIM has already achieved a scale of 100,000 spins, we contend that our research introduces a promising new quantum-inspired computational tool for expeditious traffic optimization, thus unveiling novel application scenarios for quantum optical computing.

Building upon the findings of our investigation, future research endeavors can concentrate on dimensionality reduction of high-order cost elements and achieving a more realistic approximation of real-world travel networks. This can involve varying traffic capacities based on the grade of the road sections and the heterogeneity of travelers. Moreover, the objective function devised in this study is amenable not only to the solution within experimental frameworks such as Ising machine but also to attaining optimal solution via ground-state coding using quantum algorithms like quantum annealing and quantum approximate optimization. Addressing these avenues holds promise for future quantum-inspired optimal research endeavors.

\textbf{Acknowledgements} \par
This work is supported in part by the National Natural Science Foundation of China under Grant No. 62471058, and in part by the Fund of State Key Laboratory of Information Photonics and Optical Communications, Beijing University of Posts and Telecommunications, China (No. IPOC2022ZT10)

\bibliographystyle{MSP}
\bibliography{refs}

\end{document}